\documentclass[pdflatex,sn-mathphys-num]{sn-jnl}

\usepackage{graphicx}%
\usepackage{multirow}%
\usepackage{amsmath,amssymb,amsfonts}%
\usepackage{amsthm}%
\usepackage{mathrsfs}%
\usepackage[title]{appendix}%
\usepackage{xcolor}%
\usepackage{textcomp}%
\usepackage{manyfoot}%
\usepackage{booktabs}%
\usepackage{algorithm}%
\usepackage{algorithmicx}%
\usepackage{algpseudocode}%
\usepackage{listings}%

\theoremstyle{thmstyleone}%
\theoremstyle{thmstyletwo}%

\theoremstyle{thmstylethree}%

\begin{document}

\title[Article Title]{Does a wormhole survive a cosmological bounce?}


\author*[1]{\fnm{Daniela} \sur{P\'erez}}\email{danielaperez@iar.unlp.edu.ar}

\author[1,2]{\fnm{Gustavo E.} \sur{Romero}}\email{romero@iar-conicet.gov.ar}
\equalcont{These authors contributed equally to this work.}

\author[3]{\fnm{Santiago E. } \sur{Perez Bergliaffa}}\email{sepbergliaffa@gmail.com}
\equalcont{These authors contributed equally to this work.}

\affil*[1]{\orgname{Instituto Argentino de Radioastronomía (IAR, CONICET/CIC/UNLP)}, \orgaddress{\street{Camino Gral. Belgrano Km 40}, \city{Villa Elisa}, \postcode{C.C.5, (1894)}, \state{Buenos Aires}, \country{Argentina}}}

\affil[2]{\orgname{Facultad de Ciencias Astronómicas y Geofísicas, Universidad Nacional de La Plata},  \orgaddress{\street{Paseo del Bosque s/n}, \city{La Plata}, \postcode{1900}, \state{Buenos Aires}, \country{Argentina}}}

\affil[3]{\orgdiv{Departamento de Física Teórica, Instituto de Física}, \orgname{Universidade do Estado de Rio de Janeiro}, \city{Rio de Janeiro}, \postcode{CEP 20550-013}, \state{Rio de Janeiro}, \country{Brasil}}


\abstract{We investigate whether a dynamical wormhole could survive in a universe that undergoes a cosmological bounce. First, the conditions under which a wormhole could persist from a contracting to an expanding phase of the cosmos are presented.  Then, the only two known cosmological solutions of Einstein's equations representing wormholes are analyzed, and it is shown that both dynamical wormholes exist for all cosmic times on both sides of a bouncing universe and at the bounce itself. We also provide a detailed analysis of the causal structure of such spacetimes and the matter content of the wormhole. Finally, some possible astrophysical manifestations of surviving wormholes in a bouncing universe are mentioned. Our results show that, at least for the Kim and Pérez-Raia Neto solutions, there is no topology change in the chosen cosmological model with a bounce.}




\maketitle

\section{Introduction}
\label{introduction}

 The standard cosmological model, called the $\Lambda$-CDM model, is a 6-parameter model that includes Einstein's gravity with the cosmological constant, baryons, neutrinos, photons, cold dark matter, and a hot initial state \citep{Lopez-Corredoira2017FoPh}.
 It also assumes isotropy and homogeneity of spacetime (i.e. the Friedmann-Lemaître-Robertson-Walker --FLRW-- metric). 
 The time evolution of the system is obtained by assuming the validity of nuclear and particle physics together with the Friedmann equations. The parameters of the model can be obtained from precise observations of several phenomena, such as the Cosmic Microwave Background (CMB) radiation \citep{Tristram2024A&A} and supernovae Ia \citep{Lu2022ApJ}, among others. However, the model cannot explain the apparent flatness of the universe, the horizon problem (\textit{i.e.} the reason why the CMB is so smooth and isotropic on large scales), or the origin of the density fluctuations necessary for the formation of the observed structures. Moreover, the model is singular: it fails to represent spacetime at the beginning of the cosmic expansion; all geodesics are incomplete there. 

To mitigate some of these problems, it is common to assume an inflationary period for the first moments after the onset of cosmic expansion. Inflation is a hypothetical exponential expansion of the universe at the end of the Grand Unification Epoch, $\sim 10^{-36}$ seconds after the Big Bang. In its simplest realization, inflation is driven by a scalar field subject to a conveniently chosen potential. Since inflation leads to a huge growth of the universe, it can account for both its apparent flatness and its homogeneity. In the first case, the expansion would greatly smooth out any initial curvature to near flatness. And, in the second case, since the universe before the inflationary phase was much smaller and would then have been contained in a causally connected region, it was in thermal equilibrium. In addition, the quantum fluctuations of the scalar field that drive inflation are ``frozen" during the exponential expansion, resulting in post-inflation density fluctuations that could have seeded all the observed large-scale structure. However, these features come at a cost \citep{Earman1999L,Brandenberger2008LNP,Ijjas+2013PhLB,Ijjas+2014PhLB}. What is the ``inflaton" that drives exponential growth? Can its properties be accommodated by a well-formulated fundamental theory of particle physics? Inflation also seems to require a fine-tuning of the initial conditions \citep{Earman1999L,brandenburg2005astrophysical,McCoy2015SHPMP}. Last but not least, inflation does not solve the problem of the initial singularity.

These issues led to a renovated interest in bouncing cosmologies \citep{Novello2008,Falciano:2008gt,Pinto-Neto2011,Pinto-Neto2014AN,Pinto-Neto2021,Ijjas:2016tpn}. In a bouncing scenario, the universe had a contraction phase before the usual expansion phase, and thus went through a bounce. Bouncing models are such that the initial conditions are no longer given in a very small and tiny region, perhaps of Planckian size, but in a very large and almost flat spacetime region. Thus, bouncing models not only solve the singularity problem by construction, but 
they may also be able to solve other important puzzles of the
standard cosmological model without the need for (or in conjunction with) an inflationary phase. The bounce itself requires a violation of the energy conditions, which can be implemented in several ways, either classically 
\citep{Ijjas2016}, or in a quantum scenario
\citep{Pinto-Neto2021}.

Any structure that might have been formed during the contraction phase is expected to be destroyed before the bounce,  
thus creating the conditions of the expanding universe described by the $\Lambda CDM$ model \citep{Ijjas2024JCAP}. However, it has been shown that  black holes can survive the bounce \citep{Perez2021PhRvDP,Perez2022PhRvD,Corman2022JCAP}. Dynamical black hole horizons evolve with cosmic dynamics, allowing a wide range of black holes of different sizes to survive the bounce. Such black holes, once in the expanding branch of the universe, could play a role in structure formation and gravitational wave background generation \citep{Perez2022PhRvD}. 

Another type of object that might survive a bounce is a wormhole. Unlike black holes, there is no evidence for the existence of natural wormholes, although the potential astrophysical effects caused by them have been studied extensively \citep{Torres1998PhRvD,Safonova2001PhRvD,Bambi2021Univ,Combi2024PhRvD}. A wormhole is a region of spacetime with non-trivial topology \citep{Visser1996book}. Its most distinctive feature is the so-called ``throat'', a shortcut that connects different regions of spacetime. To remain open, a throat should be maintained by exotic matter that violates the null energy condition. A large number of wormhole solutions exist in General Relativity \citep{Lobo2007}. However, in order to study the survival of a hypothetical wormhole through a cosmic bounce, a geometry describing a wormhole embedded in a suitable cosmological background should be used (for a recent review on dynamical wormholes on cosmological backgrounds, see \cite{kor+25}). 

%

In the next section, a definition of a wormhole throat that can be used in a cosmological setting is presented, and the only two known 
dynamical wormholes 
in General Relativity
immersed in a cosmological spacetime are presented. We then analyze the survival of the wormholes and discuss the implications. We close with some conclusions and perspectives.  

\section{Wormholes in a cosmological background}

Most wormholes discussed in the literature are immersed in asymptotically flat spacetimes. A wormhole is usually defined by the following conditions: (i) the associated spacetime is free of event horizons, and (ii) there is a \textit{throat} that satisfies the \textit{flare-out} condition \citep{mor+88}. The throat is the boundary that connects two separate regions of spacetime, and the flare-out conditions determine the ``shape'' of the outward curvature.

When a specific wormhole geometry is imposed on an asymptotically flat spacetime, it is found that the matter at the throat must violate the null energy condition \citep{mor+88,Visser1996book,hoc+98a}. This is perhaps the most peculiar feature of wormholes, valid even for dynamic wormholes \citep{hoc+98a}. 

Non-asymptotically flat wormholes have been studied by a number of authors \citep{mae+09,Kirilov2016IJMPD}. These solutions are usually referred to as ``cosmological wormholes''. They are constructed by connecting cosmological solutions by the wormhole's throat, and constitute actually cosmological wormhole models. Another cosmological wormhole metric was proposed by Kar and Sahdev \citep{kar94,kar+06} (see also Ref. \citep{kor+25} for further analysis on this metric). In this model, the spacetime is essentially constructed by applying a time-dependent conformal factor to the Morris–Thorne line element.

A very different type of wormhole solution of Einstein's field equations is that of a wormhole embedded in a cosmological background spacetime. Such solutions are reminiscent of the McVittie solution for a point mass in a background spacetime, and represent an astrophysical wormhole that evolves with the background cosmological model. Only two such solutions are known, the Kim solution \citep{kim96,kim18,Kim2020IJMPD} and the Pérez-Raia Neto solution \citep{per+23}. Both are asymptotically FLRW. We will introduce them in the following and investigate afterwards their behavior within a bouncing cosmological model.

\subsection{Kim cosmological wormhole solution}

The line element of this wormhole geometry is \citep{kim96}
\begin{equation}\label{kim1}
ds^{2} = - e^{2\Phi(r)} dt^{2} + R^{2}(t) \left[\frac{dr^{2}}{1 - kr^2 - b(r)/r} + r^2 d\Omega^{2}\right].
\end{equation}
Here, $\Phi(r)$ and $b(r)$ are the lapse and shape functions, respectively. As usual, $R(t)$ stands for the scale factor of the universe and $k$ is the spatial curvature. When both $\Phi(r), b(r) \rightarrow 0$, the FLRW line element is recovered. On the other hand, if we set $R(t) = 1$ and $k = 0$, the metric corresponds to the static Morris-Thorne wormhole \cite{mor+88}.

The lapse and shape functions of Kim's wormhole are respectively given by $e^{2\Phi(r)} = 1$ and $b(r) = b^2_0/r$, and a flat cosmological model, \textit{i.e.} $k = 0$ is assumed. Under these hypothesis, the line element \eqref{kim1} reads
\begin{equation}\label{kimr}
ds^{2} = -  dt^{2} + a^{2}(t) \left[\frac{dr^{2}}{1 - b^2_0/r^2} + r^2 d\Omega^{2}\right],
\end{equation}
where the normalized scale factor $a(t)$ was
introduced. The radial coordinate is defined in the range $b_0 < r < \infty$. This cosmological wormhole metric is written in coordinates $(t,r,\theta,\phi)$, where $t$ is cosmic time and $r$ is a radial coordinate. 

Under the coordinate transformation
$r \rightarrow \tilde{r}$, with  
\begin{eqnarray}
r & = & \tilde{r} \left(1 + \frac{b^2_0}{4 \tilde{r}^2}\right),\\
dr & = & \left(1 - \frac{b^2_0}{4 \tilde{r}^2}\right) \; d\tilde{r},
\end{eqnarray}
the line element \eqref{kim1} takes the form
\begin{equation}\label{kimriso}
ds^2 = -  dt^{2} + a^2(t) \left(1 + \frac{b^2_0}{4 \tilde{r}^2}\right)^2 \left[d\tilde{r}^2 + \tilde{r}^2 d\Omega^2\right].
\end{equation}

The coordinate $\tilde{r}$ is usually called isotropic radial coordinate and is defined in the interval $b_0/2 < \tilde{r} < \infty$. This cosmological wormhole metric in isotropic coordinates was analyzed by Kim in two recent papers \citep{kim18,kim20}. In Section \ref{analysis} we will explicitly compute the location of the throat and compare with the results given by Kim in 
\citep{kim18,kim20}.

The density $\rho$ and radial $p_{\mathrm r}$ and tangential $p_{\mathrm t}$ pressures associated to the energy-momentum tensor which sources Kim's wormhole are
\begin{eqnarray}
\rho_{\mathrm{K}} & = & -\frac{32}{\pi}\frac{b^2_0 \tilde{r}^4}{a^2(t)\left(b^2_0 + 4 \tilde{r}^2\right)^4},\label{rhoK}\\
p_{\mathrm{rK}} & = & - \frac{32}{\pi}\frac{b^2_0 \tilde{r}^4}{a^2(t)\left(b^2_0 + 4 \tilde{r}^2\right)^4},\label{prk}\\
p_{\mathrm{tK}} & = &  \frac{32}{\pi}\frac{b^2_0 \tilde{r}^4}{a^2(t) \left(b^2_0 + 4 \tilde{r}^2\right)^4},\label{ptk}
\end{eqnarray}
where the subscript ``K'' refers to Kim's solution.




\subsection{Pérez-Raia Neto cosmological wormhole solution}

The general form of the line element of the cosmological wormhole spacetime proposed by  \cite{per+23} in isotropic coordinates $(t,\tilde{r},\theta,\phi)$ takes the form \footnote{The redshift function is $\Phi'=0$ as to ensures the absence of event horizons.}
\begin{equation}\label{gcw}
ds^{2} = -dt^2 + a^2(t)\left(1+\frac{b(\tilde{r},t)^2}{4 \tilde{r}}\right)^2 \left(d\tilde{r}^2+\tilde{r}^2 d\Omega^2\right).
\end{equation}
Here, $b(\tilde{r},t) = b_1(\tilde{r}) \times b_2(t)$, where $b_2(t)$ is an arbitrary function of the cosmic time. The radial part of the shape function was chosen as
\begin{equation}
b_1(\tilde{r}) = \frac{b_0}{\sqrt{\tilde{r}}}.  
\end{equation}

Note that in the limit $a(t) \rightarrow 1$, and setting $b_{1}(\tilde{r}) = b_0/\sqrt{\tilde{r}}$ and $b_{2}(t) = 1$, the line element of the Morris-Thorne wormhole \citep{mor+88} is recovered; by choosing $b_{1}(\tilde{r}) = b_0/\sqrt{\tilde{r}}$ and $b_{2}(t) = 1$, the metric corresponds to Kim's cosmological wormhole. 

We focus on a specific model for the wormhole setting $b_2(t) \equiv 1/a(t)$. Thus, the line element reads
\begin{equation}\label{23}
ds^{2} = -dt^2 + a^2(t)\left(1+\frac{b^2_0}{4 a^{2}(t)\tilde{r}^2}\right)^2 \left(d\tilde{r}^2+\tilde{r}^2 d\Omega^2\right).   
\end{equation}

In \cite{per+23} it was shown that the throat is located at $\tilde{r} = b_0/(2 a(t))$. In these coordinates the location of the throat evolves 
following
the dynamics of the cosmological background.

The energy-momentum tensor that corresponds to this solution is that of an imperfect fluid. The expressions for the density $\rho_{\mathrm{PRN}}$,  radial $p_{r\mathrm{PRN}}$ and tangential pressures  $p_{t\mathrm{PRN}}$ and the heat flux $q$ are \citep{per+23}
\begin{eqnarray}
\rho_{\mathrm{PRN}} & = - & \frac{2 b^2_0 \tilde{r}^2 \tilde{\rho}(\tilde{r},t)}{\pi \beta^4},\label{den-us}\\
p_{r\mathrm{PRN}} & = & \frac{b^2_0  \tilde{pr}(\tilde{r},t)}{2 \pi a^2(t) \beta^4},\label{pr-us}\\
p_{t\mathrm{PRN}} & = & \frac{b^2_0  \tilde{pt}(\tilde{r},t)}{2 \pi a^2(t) \beta^4},\label{pt-us}\\
q & = & \frac{64 b^2_0 \; \tilde{r}^5 a'(t) \; a^3(t)}{\pi \beta^4}\label{q},
\end{eqnarray}
where
\begin{eqnarray}
\tilde{\rho}(r,t) & = & 16 \tilde{r}^2 a^4(t) + 3 \beta^2 a'(t)^2,\\
\tilde{p}_{r}(r,t) & = & -64 \tilde{r}^4 a^6(t) + \beta^2  \chi,\\
\tilde{p}_{t}(\tilde{r},t) & = &64 \tilde{r}^4 a^6(t) +\beta^2  \chi,\\
\beta & = & 4 a^2 (t) \tilde{r}^2 + b^2_0,\\
\chi & = & - b^2_0 a'^2(t) + a(t) \beta a''(t).
\end{eqnarray}
The subscript ``PRN'' refers to the Pérez Raia-Neto cosmological wormhole.

\section{Evolution of the throat through a cosmological bounce}

Since the existence of a throat and the fulfillment of the flaring-out condition are the defining characteristics of wormhole spacetimes, we need to examine whether these features are maintained at every cosmic epoch for the Kim and Pérez–Raia Neto models.

In static spacetimes, the location of the wormhole throat can be easily identified by inspecting the spatial part of the line element. It coincides with the wormhole's center, and is determined by solving the implicit equation $b(r_0) = r_0$, where $b(r)$ is the shape function characterizing the wormhole geometry. However, as discussed by Hochberg and Visser \citep{hoc+98b}, this procedure does not generalize to dynamical wormhole spacetimes. In such cases, the identification of the throat requires a more rigorous approach. In particular, we need a precise definition of a throat and a criterion for its existence once the throat is conveniently defined in a given dynamical spacetime. This is detailed below, following the formal definitions established by Hochberg and Visser.

\subsection{Definition of the throat.}\label{def-thr}

Instead of using a definition such as the one provided by the classical paper by Morris and Thorne \citep{mor+88}, here we adopt a more general approach
that includes dynamical wormholes.
Following \cite{hoc+98b},
the throat
is a minimal two-surface where null rays coming from one side are converging, and start to diverge on the other side
\footnote{A similar definition was provided by \cite{Tomikawa2015PhRvD}, see also the paper by \cite{Hayward1999IJMPD}.}. The throat can connect two different regions of the same cosmic hypersurface or regions from different hypersurfaces.

If we denote by $n^{a}$ and $l^{a}$ the ingoing and outgoing tangent fields of null radial geodesics, and by $\theta_{n}$ and $\theta_{l}$ their respective expansions, one of the following conditions is satisfied on the throat \citep{mcn+21}
\begin{equation}
\theta_{n} = 0 \;\; \wedge \;\; n^{a}\nabla_{a}\theta_{n} \ge 0, \label{throat1}
\end{equation}
or,
\begin{equation}
\theta_{l} = 0 \;\; \wedge \;\; l^{a}\nabla_{a}\theta_{l} \ge 0. \label{throat2}
\end{equation}

The condition $n^{a}\nabla_{a}\theta_{n} \ge 0$ ($l^{a}\nabla_{a}\theta_{l} \ge 0$) is the generalization of the Morris-Thorne flare-out condition for dynamical wormholes. 

\subsection{Survival condition}\label{sec:survival_condition}

Let assume that we have a bouncing spacetime $(M, {\bf g})$ such that a sequence of evolving spacelike ``instantaneous" hypersurfaces $S_1, S_2,...,S_n$
can be defined on it, with $S_{i+1}\subset I^+(S_i)$, where $I^+(S_i)$ is the the causal future of $S_i$ \cite{Joshi1987PhLA}. Let us now consider that the hypersurface $S'\subset I^-(S)$ belongs to the contracting branch of spacetime, and $S''\subset I^+(S)$ belongs to the expanding branch, with $S$ the minimal surface at the bounce, where the scale factor of the spacetime is such that $\dot{a}=0$. Then, if there is a wormhole throat $T_{S'}$ in $S'$, all spacelike hypersurfaces between $S'$ and $S''$ should be diffeomorphic. If this is not the case, there should be a topology change between $S'$ and $S''$ and the throat must close. 

This is a special case of a general theorem of \cite{Borde1994}, which states that a spacetime cannot display a topology change if it is hole-free (i.e., causally compact) and can be foliated by spacelike hypersurfaces.

To determine whether there are topological changes associated with the destruction of the wormhole embedded in the bouncing spacetimes to be discussed, we will parameterize the scalar throat conditions \eqref{throat1} or \eqref{throat2} in terms of cosmic time, and we will examine whether they hold for all cosmic times on both sides of the bounce and at $S(t_{\rm bounce})$. 


\section{Analysis}\label{analysis}

\subsection{Bouncing model}

In a bouncing cosmology, the universe contracts from a
very dilute phase. The contraction then smoothly evolves
into a bounce that leads to the current phase of expansion as 
described by the $\Lambda$CDM model. By construction, the
cosmological singularity is absent in such bouncing models.
There are many mechanisms that could cause a
cosmological bounce, either by classical \citep{Ijjas2016,Galkina:2019pir} or quantum \citep{Peter:2008qz,Almeida:2018xvj,Bacalhau:2017hja} effects. The reader is referred to the review 
by \citep{Novello2008} for further details on these mechanisms.

Here we take a phenomenological view by characterizing the model with a few free parameters, rather than restricting ourselves to specific bouncing theories. Specifically, we will use the following parameterization of the scale factor 
\cite{Agullo+2021}:
\begin{equation}
    a(t)= a_{b} (1+\xi t^2)^n,
\end{equation}
with constants $\xi$ and $n$. It is easy to check that $\xi$ determines the value of the Ricci curvature scalar at the bounce, namely $R_{\rm b} = 12 \xi n$. Thus, a bounce in this family of models is characterized by two parameters, $n$ and $R_{\rm b}$, which encode the new physics causing the bounce. To make specific calculations, we take $n = 1/3$
\cite{pet+07}:
\begin{equation}
    a(t)=a_{\rm b}\left[1+ \left(\frac{t}{T_{\rm b}}\right)^2\right]^{1/3}, \label{a-bounce}
\end{equation}
where $T^2_{\rm b} = 1/ \xi$. The bounce occurs when the curvature of spacetime becomes very large. At $t>>T_{\rm b}$ the scale factor reduces to that of dust. Near the bounce the evolution is driven by
by an effective fluid with negative energy density, which scales as $a^{-6}$, as can be seen from Friedmann's equation.
The violation of the energy conditions around $T_{\rm b}$ ensures the non-applicability of the singularity theorems \cite{Hawking-Ellis-book}, making spacetime regular at all cosmic times.


\subsection{Kim wormhole}

The location of the wormhole throat is calculated using the definition provided in Section \ref{def-thr}, by means of the 
areal radius coordinate $R$
\begin{equation}\label{areal-kim}
R = a(t) \left(1 + \frac{b^2_0}{4\tilde{r}^2}\right) \tilde{r},
\end{equation}
in terms of which the line element reads
\begin{eqnarray}
ds^2 & = & -dt^2 \left[1 - \frac{R^2 H^2}{\left(1 - \frac{b^2_0 a^2(t)}{R^2}\right)}\right] + \frac{dR^2}{\left(1 - \frac{b^2_0 a^2(t)}{R^2}\right)}\nonumber \\ 
& - & \frac{2 R H dR dt}{\left(1 - \frac{b^2_0 a^2(t)}{R^2}\right)} + R^2 d\Omega^2.    
\end{eqnarray}
Here, $H \equiv \dot a/a$ is the Hubble factor.

Since the metric is spherically symmetric, the ingoing and outgoing null radial geodesics 
can be easily obtained
by setting $\theta = \pi/ 2$ and $d\phi = 0$ in $ds^2 = 0$. We get
\begin{equation}\label{geo_kim}
\left. \frac{dR}{dt}\right\vert_{\pm} = RH \pm  \sqrt{1- \frac{b^2_0 a^2(t)}{R^2}},
\end{equation}
where the ``$ - $'' (``$+$'') corresponds to ingoing (outgoing) case. The tangent vector fields $n^{a}$ and $l^{a}$ have the form
\begin{eqnarray}
n^{\mu} & = & \left(1,R H - \sqrt{1- \frac{b^2_0 a^2(t)}{R^2}},0,0\right),\label{na}\\
l^{\mu} & = & \left(1,R H + \sqrt{1- \frac{b^2_0 a^2(t)}{R^2}},0,0\right).\label{la}
\end{eqnarray}

The corresponding expansions $\theta_{n}$ and $\theta_{l}$ are determined using an expression that takes into account the case in which the geodesic 
is not necessarily affinely-parametrized, namely \cite{far15}
\begin{eqnarray}
\theta_{n} & = & \left[ g^{ab} + \frac{l^{a} n^{b} + n^{a} l^{b}}{- n^{c} l^{d} g_{cd}}\right] \nabla_{a} n_{b}, \\
\theta_{l} & = & \left[ g^{ab} + \frac{l^{a} n^{b} + n^{a} l^{b}}{- n^{c} l^{d} g_{cd}}\right] \nabla_{a} l_{b}.
\end{eqnarray}
Inserting \eqref{na} and \eqref{la} into the two formulae above, we obtain
\begin{eqnarray}
\theta_{n} & = & -2 \frac{\sqrt{1- \frac{b^2_0 a^2(t)}{R^2}}}{R}+ 2 H,\\
\theta_{l} & = & 2 \frac{\sqrt{1- \frac{b^2_0 a^2(t)}{R^2}}}{R}+ 2 H.
\end{eqnarray}
The first equality in condition \eqref{throat1} is satisfied if 
\begin{equation}\label{sol-kim}
R_{\pm} = \frac{\sqrt{1 \pm \sqrt{1 - 4 b^2_0 H^2 a^2(t)}}}{\sqrt{2} \lvert H \rvert}. 
\end{equation}

The same result is obtained from the first equality in condition  \eqref{throat2}\footnote{The location of the throat in isotropic coordinates follows from inserting Eq. \eqref{sol-kim} ($R_{-})$ into
\begin{equation}
\tilde{r} = \frac{R - \sqrt{R^2 - b^2_0 a^2(t)}}{2a(t)}.
\end{equation}

This is not clearly stated in Kim's papers and the reader might mistakenly think that $\tilde{r} = b_0/2$ corresponds to the radius of the throat, which is not the case.
}. We have discarded those solutions yielding negative values of $R$. Kim arrived at the same result in his two works: the throat corresponds to $R_{-}$ and $R_{+}$ to the cosmological horizon. The throat is only defined if $1 - 4 b^2_0 H^2 a^2(t) > 0$. 

Let us associate a mass with the parameter $b_0$, the latter with units of length $b_0 (M) = G M/c^2$, and define the function $f_{\rm{c}}$ as follows:
\begin{equation}\label{fc}
f_{\rm{c}}(t,M) = 1 - 4 b^2_0 H^2 a^2(t).
\end{equation}

\begin{figure}
    \centering
    \includegraphics[width=1.\linewidth]{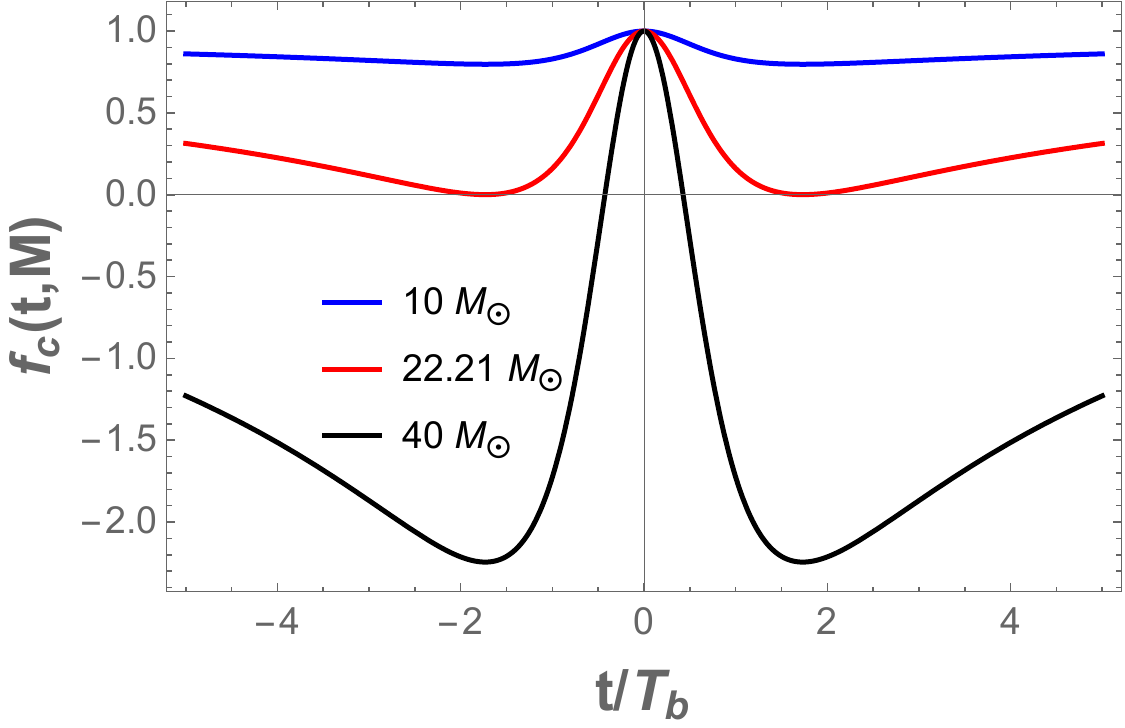}
    \caption{Plot of expression \eqref{fc} as a function of the cosmic time for $b_0 = 10, 22.21, 40 \; G M_{\odot} /c^2$.}
    \label{fig_0_a}
\end{figure}

In Fig. \ref{fig_0_a} we show a plot of $f_{\rm{c}}$ as a function of the cosmic time for different values of the throat mass: $b_0 = 10 \; M_{\odot}$, $ b_0 = 22.21 \; M_{\odot}$ and $b_0 = 40 \; M_{\odot}$. It is seen that $f_{\rm{c}}$ has two minima which are symmetrical with respect to $t = 0$. By calculating $df_{\rm{c}}(t,M)/dt = 0$ we find that these minima occur at $t/T_{\rm B} = \pm 1.73$. Then, the maximum value of $b_0$ such that $f_{\rm{c}} \ge 0$ is $b_0 = 22.21 \; M_{\odot}$. This calculation was done by adopting the scale factor \eqref{a-bounce} and taking $T_{b} = 10^{-4}$ s \citep{Frion2020}. So, condition $f_{\rm{c}} \ge 0$ is only satisfied at all cosmic times for $b_{0} \in [0,G M^{*}/c^2]$, with $ M^{*} \approx  22 \; M_{\odot}$.

The compliance with the generalized flare-out was not checked in Kim's papers \citep{kim18,kim20}. This is done next:
\begin{eqnarray}
\left.n^{a}\nabla_{a}\theta_{n}\right\vert_{\mathrm{thr}}  & = & \frac{4 \left(f_1 +f_2 + f_3\right)}{\left(1 - \sqrt{1 -4 b^2_0 a^2(t) H^2(t)}\right)^2} ,\label{flaring-out-cond-kim}\\
f_1 & = & 6 b^2_0 \; a(t) \; H^3 a'(t),\\
f_2 & = & H' \left(1 - \sqrt{1 -4 b^2_0 a^2(t) H^2(t)}\right),\\
f_3 & = & - b^2_0 a^2(t) H^2 \left(2 H^2 + H'\right).
\end{eqnarray}

\begin{figure}
    \centering
    \includegraphics[width=1.\linewidth]{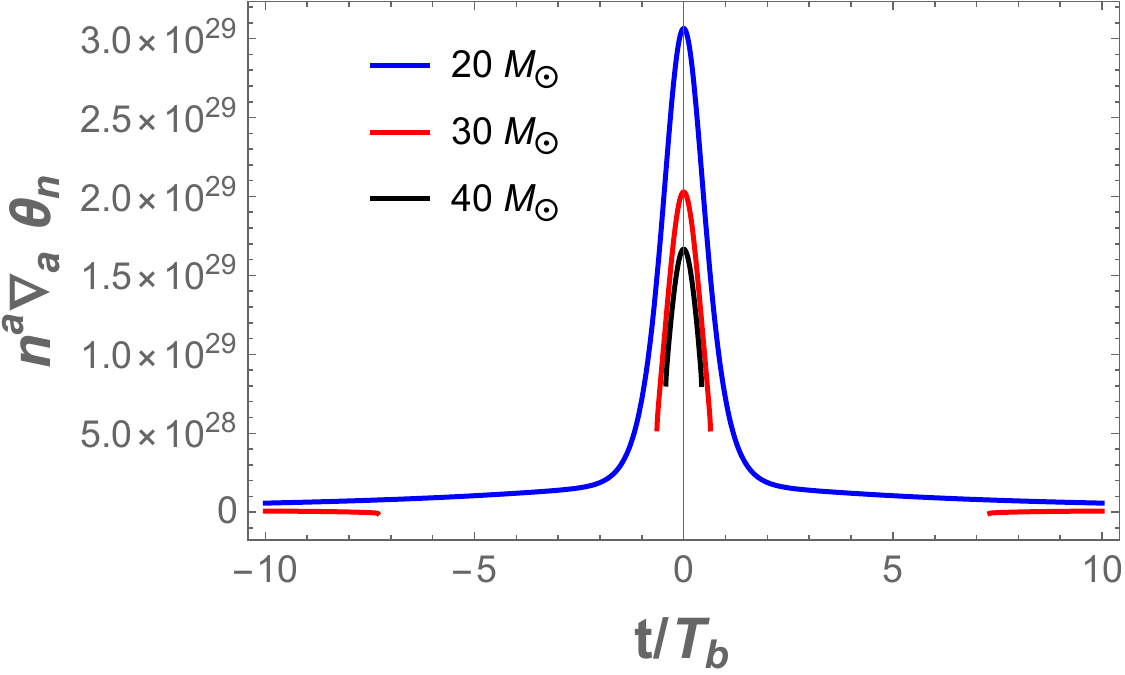}
    \caption{Plot of expression \eqref{flaring-out-cond-kim} as a function of the cosmic time for $b_0 = 20, 30, 40 \; G M_{\odot} /c^2$.}
    \label{fig_1}
\end{figure}

The flaring-out condition, which is satisfied if the expression \eqref{flaring-out-cond-kim} is equal to or greater than zero, is fulfilled 
at all cosmic times 
only for $b_{0} \in (0,G M^{*}/c^2)$ . Otherwise, it is negative or even complex in certain ranges of values of the cosmic time. This is shown in Figure \ref{fig_1}.

\begin{figure}
    \centering
    \includegraphics[width=1.\linewidth]{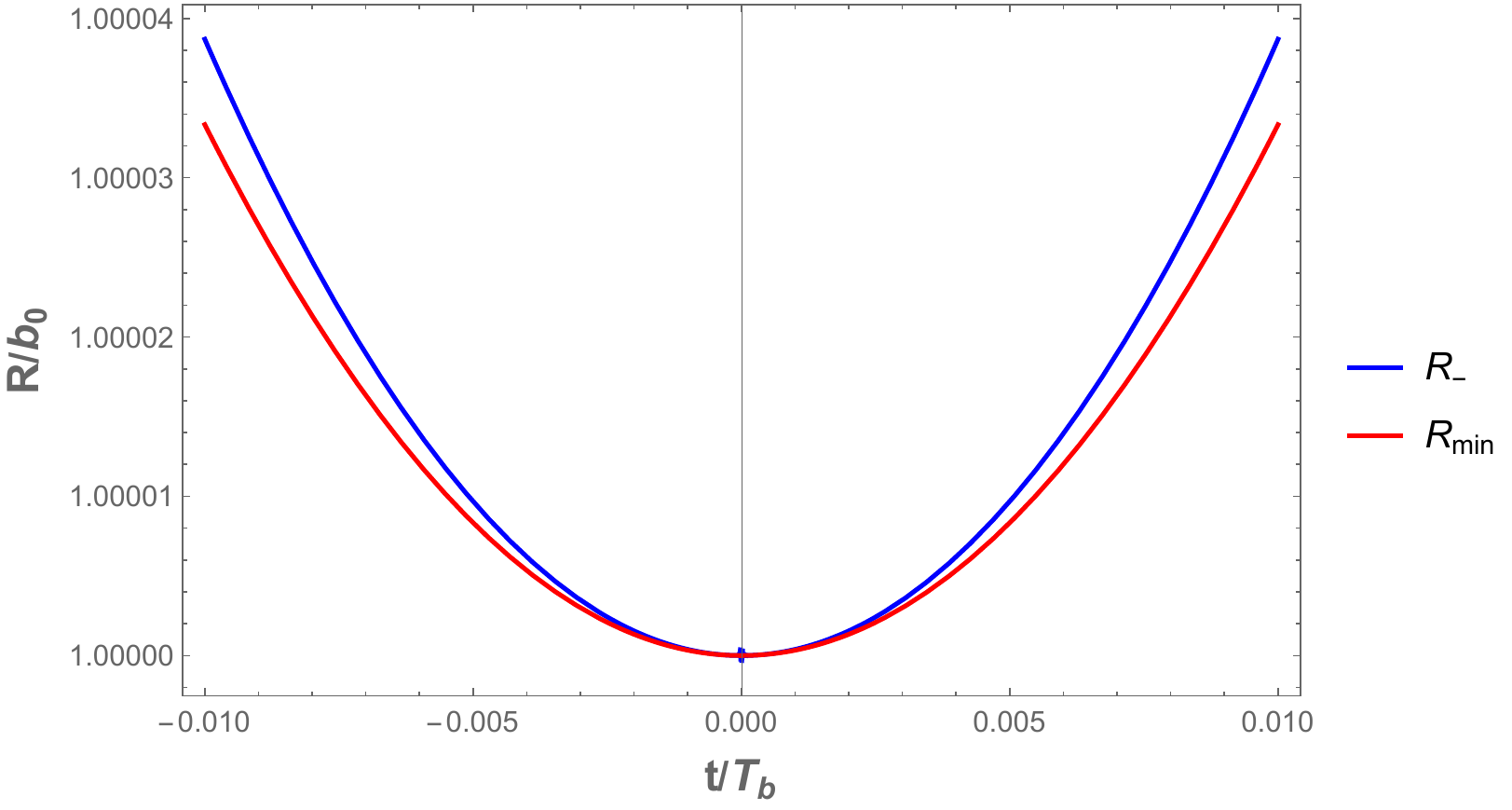}
    \caption{Plot of the areal radius of the throat for Kim wormhole as a function of the cosmic time (blue line). The red curve represents the minimum value of the areal radius coordinate.}
    \label{fig_1a}
\end{figure}

We show a plot of the throat, that is $R_{-}$, as a function of the cosmic time in Figure \ref{fig_1a}. For comparison, we also include the curve $R_{\mathrm{min}} = b_0 \; a(t)$, which represents the minimum value of the areal radius coordinate. This expression is obtained by setting $\tilde{r} = b_0/2$ in Eq. \ref{areal-kim}. We see that the throat is well-defined at all cosmic times and is always larger than $R_{\mathrm{min}}$, except at the bounce, where they coincide. 

Given the survival condition, stated in \ref{sec:survival_condition}, Kim wormhole survives the cosmological bounce adopted in this work. However, if a mass is associated with the parameter $b_0$, then the throat only exists for $b_0  \in [0,G M^{*}/c^2]$, with $M^{*} \approx  22 \; M_{\odot}$. The flare-out condition is also satisfied within this same range of  $b_0$. 

\begin{figure}
    \centering
    \includegraphics[width=1.\linewidth]{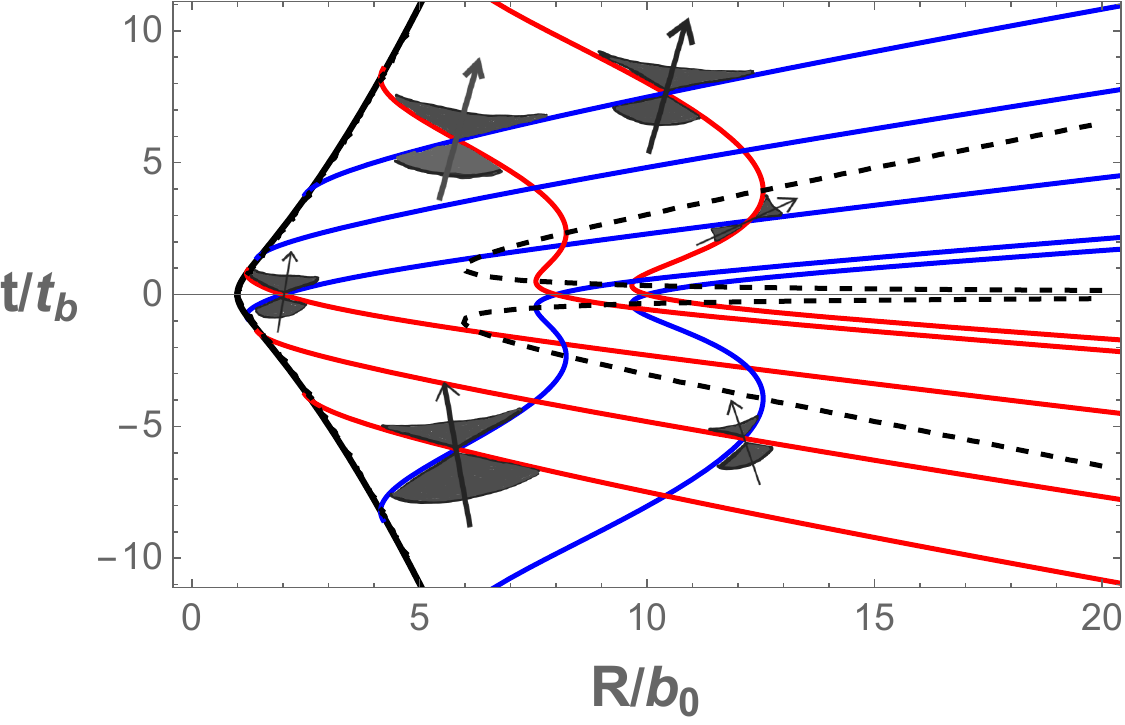}
    \caption{Plot of the causal structure of the Kim wormhole solution. The red (blue) curves represent the null ingoing (outgoing) radial geodesics. The grey shaded regions show some light cones and the black arrow indicates the future direction. The black curve corresponds to the location of the throat.  
    The dashed black curves show the points at which the condition $dR/dt = 0$ is satisfied.
    We adopt $T_b = 10^{-4}$ and $b_0 = 10 \; G M_{\odot} /c^2$. }
    \label{fig_1b}
\end{figure}

The radial ingoing (``$-$'') and outgoing (``$+$'') null geodesics, obtained by integrating Eq.~\eqref{geo_kim}, are shown in Figure~\ref{fig_1b}. The red (blue) curves represent the null ingoing (outgoing) radial geodesics. The grey shaded regions indicate selected light cones, and the black arrow denotes the direction of increasing time. The black curve marks the location of the throat, while the dotted curves correspond to the condition $dR/dt = 0$: the curve in the region $t >0$ ($t < 0$) is for the ingoing (outgoing) null geodesics.

We see that the radius of the throat (in areal radius coordinate) is minimum at the bounce. As expected, all null geodesics remain confined to the region \( R \geq R_{-} \) before, during, and after the bounce.

We conclude the analysis of Kim's solution by calculating the embedding diagram related to the two-surface given by the line element \eqref{kimriso} (we set $t = t_{*} = \mathrm{constant} $ and $\theta = \pi /2$)
\begin{equation}\label{1-cosmo}
ds^2=f(\tilde r, t_*)^2(d\tilde r^2+\tilde r^2d\phi^2)
\end{equation}
with
\begin{equation}
f(\tilde{r},t_{*}) = a(t_{*})\left(1+\frac{b^2_0}{4 \tilde{r}^2}\right) .
\end{equation}

To visualize this two-surface, we embed it in a 3-dimensional Euclidean space written in cylindrical coordinates $(\rho, \phi, z)$ as
\begin{equation}\label{3d-eucl}
ds^{2} = dz^2 + d\rho^2 + \rho^2 d\phi^2 = \left[\left(\frac{dz}{d\tilde{r}}\right)^2 + \left(\frac{d\rho}{d\tilde{r}}\right)^2\right] d\tilde{r}^2+ \rho^2 d\phi^2.
\end{equation}

\begin{figure*}
    \centering
    \includegraphics[width=1.\linewidth]{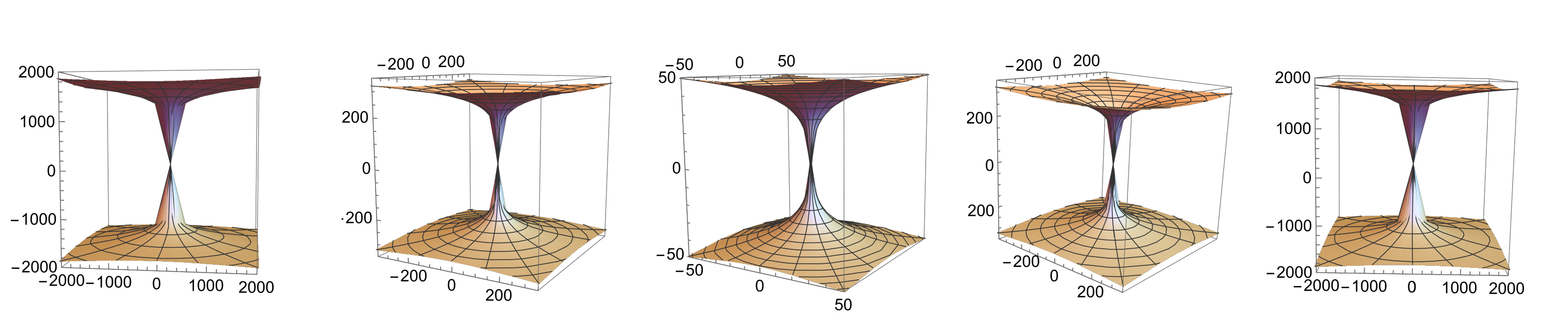}
    \caption{Embedding diagram for the Kim wormhole for $t/T_{\rm {b}} =-100, -10, 0, 10, 100$ (from left to right). We adopt $b_0 = 10 \; G M_{\odot} /c^2$. }
    \label{fig_2k}
\end{figure*}

By comparing \eqref{1-cosmo} and \eqref{3d-eucl}, we find that
\begin{eqnarray}
\frac{dz}{d\tilde{r}} & = & \pm \sqrt{a^2(t_{*})\left(1+\frac{b^2_0}{\tilde{r}^2}\right)^2 - \left(\frac{d\rho}{d\tilde{r}}\right)^2 },\nonumber \\
& = & \pm a(t_{*})\; \frac{b_0}{\tilde{r}},
\end{eqnarray}
so,
\begin{equation}
z(\tilde{r}) = \pm a(t_{*}) \; b_0 \;\log{(\tilde{r}/b_0)} + \mathrm{const}.
\end{equation}
The embedding diagram is shown in Figure~\ref{fig_2k}. The shape of the wormhole remains unchanged at all cosmic times. The entire structure, however, is either contracted or expanded by the action of the scale factor.

\subsection{Pérez-Raia Neto wormhole}

Pérez and Raia Neto explicitly calculated conditions \eqref{throat1} and \eqref{throat2} for the line element \eqref{23}.
We state here only the main results; the detailed calculations can be found in \citep{per+23}, where the line element was 
written in terms of the areal radius coordinate $R$, defined as
\begin{equation}
R = a(t) \left(1 + \frac{b^2_0}{4 a^2(t) \tilde{r}^2}\right) \tilde{r}.
\end{equation}

\begin{figure}
    \centering
    \includegraphics[width=1.\linewidth]{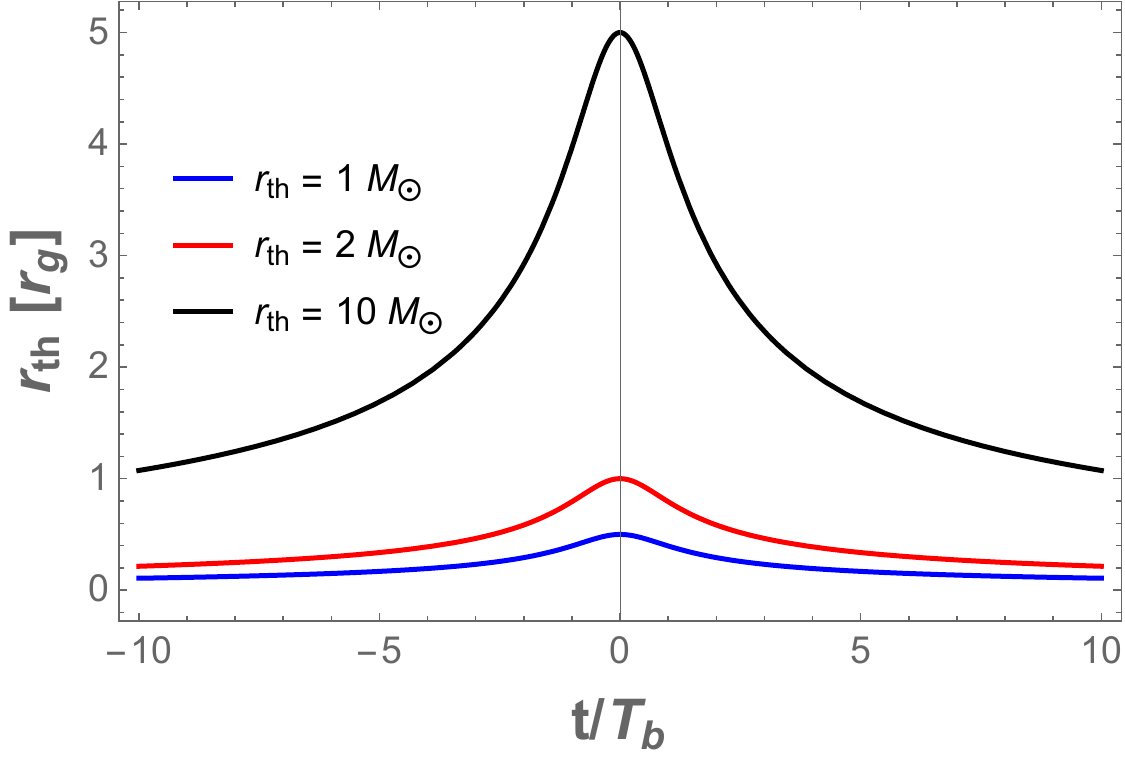}
    \caption{Radius of the throat for the PRN wormhole in isotropic coordinates as a function of the cosmic time. We adopt $b_0 = 1, 2, 10 \; G M_{\odot} /c^2$.}
    \label{fig_0}
\end{figure}

The cosmological wormhole metric in coordinates $(t, R, \theta, \phi)$ reads
\begin{eqnarray}
ds^{2} & = & - \left(1- R^2 H^2\right) dt^2 + \frac{1}{\left(1 + \frac{b^2_0}{R^2}\right)} dR^2 \nonumber \\
& - & 2\frac{R H}{\sqrt{1- \frac{b^2_0}{R^2}}} dR dt + R^2 d\Omega^2.
\end{eqnarray}
The throat is at $R = b_0$ and there is a cosmological horizon at $R = 1/H$. The location of the throat in isotropic coordinates is $\tilde{r}_{\rm th} = b_0/(2 a(t))$. We show in Fig. \ref{fig_0} a plot of $\tilde{r}_{\rm th}$ as a function of the cosmic time for different values of the parameter $b_0$. The behavior of the throat radius (in isotropic coordinates) is due to the scaling with $a^{-1}(t)$.  

The flare-out condition is also satisfied in the throat. Hence, the throat is perfectly defined for all cosmic times and, in particular, for the bouncing cosmology characterized by the scale factor \eqref{a-bounce}. Thus, this cosmological wormhole solution satisfies the previously stated survival condition, and the wormhole exists for all cosmic times on both sides of the bounce and at the bounce itself.

To better understand the causal structure of this solution before and after the bounce, we integrate the equation of the radial ingoing (``$-$'') and outgoing (``$+$'') null geodesics:
\begin{equation}
\left. \frac{dR}{dt}\right\vert_{\pm} = \left(\pm 1 + RH\right) \sqrt{1 - \frac{b^2_0}{R^2}}.
\end{equation}

\begin{figure}
    \centering
    \includegraphics[width=1.\linewidth]{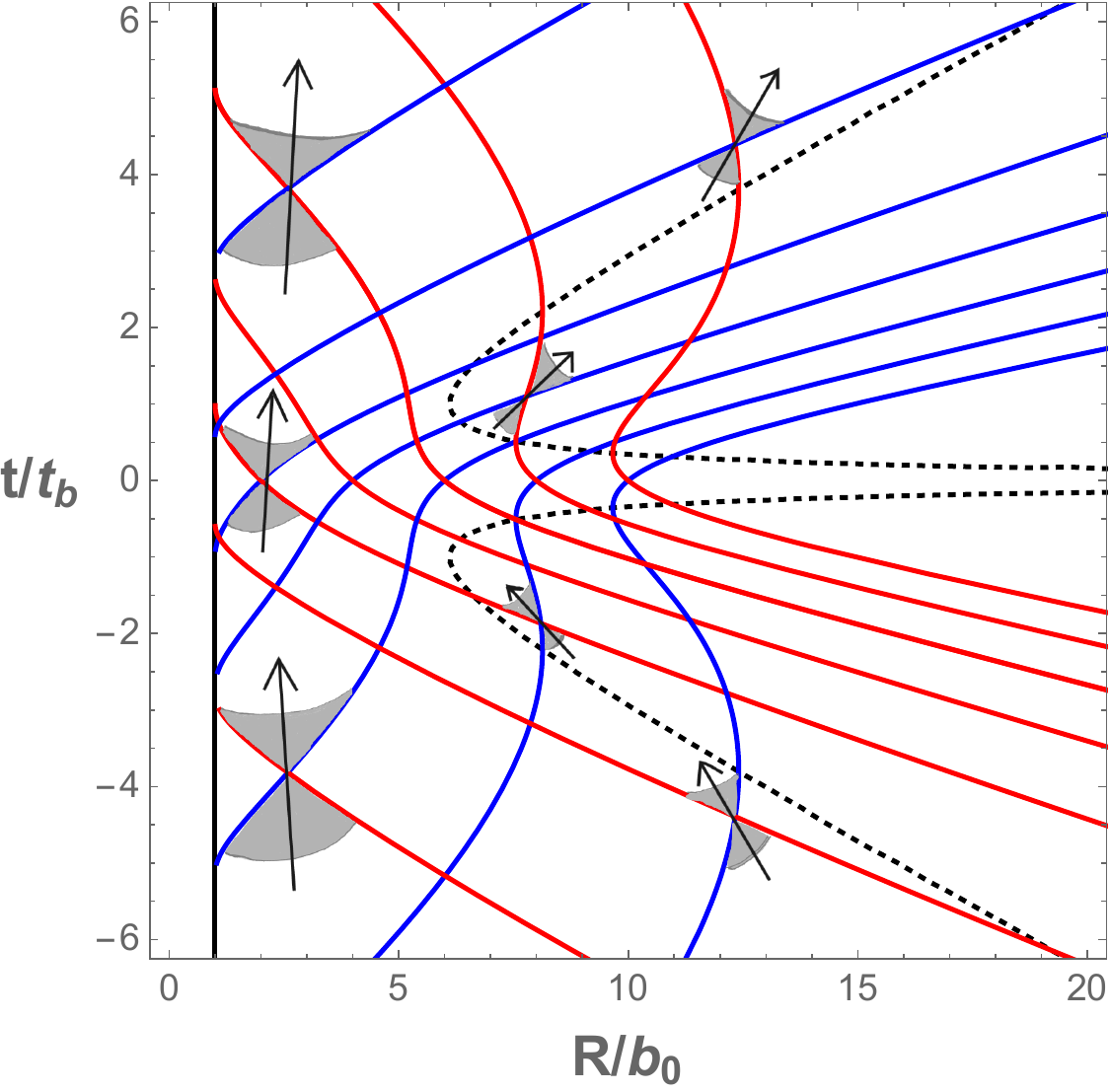}
    \caption{Plot of the causal structure of the Perez-Raia Neto wormhole solution. The red (blue) curves represent the null ingoing (outgoing) radial geodesics. The grey shaded regions show some light cones and the black arrow indicates the future direction. The black vertical line corresponds to the location of the throat. 
        The dashed black curves show the points at which the condition $dR/dt = 0$ is satisfied.
In this figure we adopt $T_b = 10^{-4}$ and $b_0 = 10 \; G M_{\odot} /c^2$. }
    \label{fig_2}
\end{figure}

The result is shown in Figure \ref{fig_2}. The red (blue) curves represent the null ingoing (outgoing) radial geodesics. The gray shaded regions show some light cones, and the black arrow indicates the future direction. The black vertical line corresponds to the location of the throat. The dashed black curves show the condition $dR/dt = 0$: the curve in the region $t > 0$ ($t < 0$) is for the ingoing (outgoing) null geodesics.

The throat of the cosmological wormhole is present at all cosmic times. There are no geodesics whose radial coordinate is less than $b_0$, as expected, since the throat is a minimal two-surface.

We also compute the embedding diagram associated with the two-surface given by the line element \eqref{23} after setting $t = t_{*} = $ constant and $\theta = \pi/2$
\begin{equation}\label{2-cosmo}
ds^2=f(\tilde r, t_*)^2(d\tilde r^2+\tilde r^2d\phi^2)
\end{equation}
with
\begin{equation}
f(\tilde{r},t_{*}) = a(t_{*})\left(1+\frac{b^2_0}{4 a^{2}(t_{*})\tilde{r}^2}\right).
\end{equation}
Following the same steps as in the previous section, the function $z(r)$ takes the form
\begin{equation}
z(\tilde{r}) = \pm b_0 \;\log{(\tilde{r}/b_0)} + \mathrm{const}.
\end{equation}

\begin{figure*}
    \centering
    \includegraphics[width=1.\linewidth]{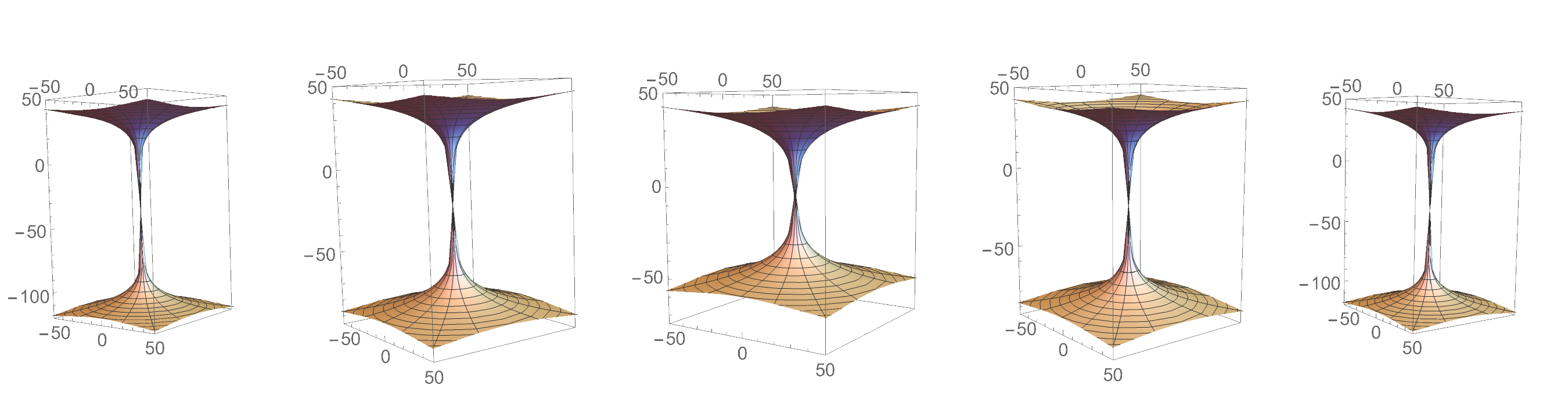}
    \caption{Embedding diagram for the Perez-Raia Neto wormhole for $t/T_{\rm {b}} =-100, -10, 0, 10, 100$ (from left to right). We adopt $b_0 = 10 \; G M_{\odot} /c^2$. }
    \label{fig_3}
\end{figure*}

In this simple result, the dependence on the scale factor has been canceled out. However, the isotropic coordinate $\tilde{r}$ is defined for $\tilde{r} \ge b_0/(2 a(t))$, so the embedding diagram of the wormhole changes as a function of cosmic time. This is shown in Figure \ref{fig_3} for $t/T_{\rm{b}} = -100, 10, 0, 10, 100$. We see that, in isotropic coordinates, the radius of the throat decreases for $t \rightarrow \infty$ ($t \rightarrow - \infty$). 

\section{Analysis of the wormhole matter content}

In the previous section we showed that the wormhole described by the Pérez-Raia Neto wormhole solution survives a cosmological bounce. Next, the main features of the matter content associated with this  wormhole are analyzed.




Let us begin by showing that the matter threading the throat violates the null-energy condition (NEC) before, after, and at the bounce\footnote{The NEC expression adopted in the present work is the correct one, in contrast to that presented in the original paper introducing the Pérez–Raia Neto solution (P\'erez, D., Neto, M.R.: A new solution for a generalized cosmological wormhole. \textit{European Physical Journal C} \textbf{83}(12), 1127 (2023)).}. For an imperfect fluid, the violation of the NEC is equivalent to that al least one the following three inequalities is satisfied \citep{mae+22}:
\begin{equation}
(\rho + p_r)^2 - 4 q^2 < 0,\;\;\; \mathrm{or} \;\;\; \rho + p_r < 0,\;\; \mathrm{or}\;\;\; \rho-p_r+2 p_t + \sqrt{(\rho + p_r)^2 - 4q^2} < 0.
\end{equation}
From Eqs. \eqref{den-us}, \eqref{pr-us}, \eqref{pt-us}, and \eqref{q}, evaluating the first two inequalities at the throat $\tilde{r}_{\rm th} = b_0/(2 a(t))$, we get
\begin{eqnarray}
(\rho + p_r)^2 - 4 q^2\vert_{\tilde{r}_{\rm th}} & = & \frac{ a^4(t)+ b^2_0 (a'(t))^2 + b^4_0 (a'(t))^4}{16 \; b^4_0 \; \pi^2 \; a^4(t)},\label{necb1}\\
\rho + p_r \vert_{\tilde{r}_{\rm th}} & = & - \frac{1}{4\pi} \frac{a^2(t) + b^2_0 a'(t)}{b^2_0 a^2(t)}\label{necb2}.
\end{eqnarray}
While Eq. \eqref{necb1} is always positive regardless of the specific form of the scale factor in the bouncing model, the opposite behavior for Eq. \eqref{necb2}. Thus, the NEC is violated at the throat for all values of the cosmic time in a bouncing universe.

For the sake of completeness, we compute the left hand side of the third inequality
\begin{equation}
\rho-p_r+2 p_t + \sqrt{(\rho + p_r)^2 - 4q^2}\vert_{\tilde{r}_{\rm th}} = \frac{a^2(t) - b^2_0 (a'(t))^2 + \sqrt{a^4(t) + b^2_0 (a'(t))^2+ b^4_0 (a'(t))^4}}{4 \pi\  b^2_0 \  a^2(t)}\label{cond3}
\end{equation}
We see that that, at least at the bounce, $a'(t) = 0$, so Eq. \eqref{cond3} remains positive.

\begin{figure}
    \centering
    \includegraphics[width=1.\linewidth]{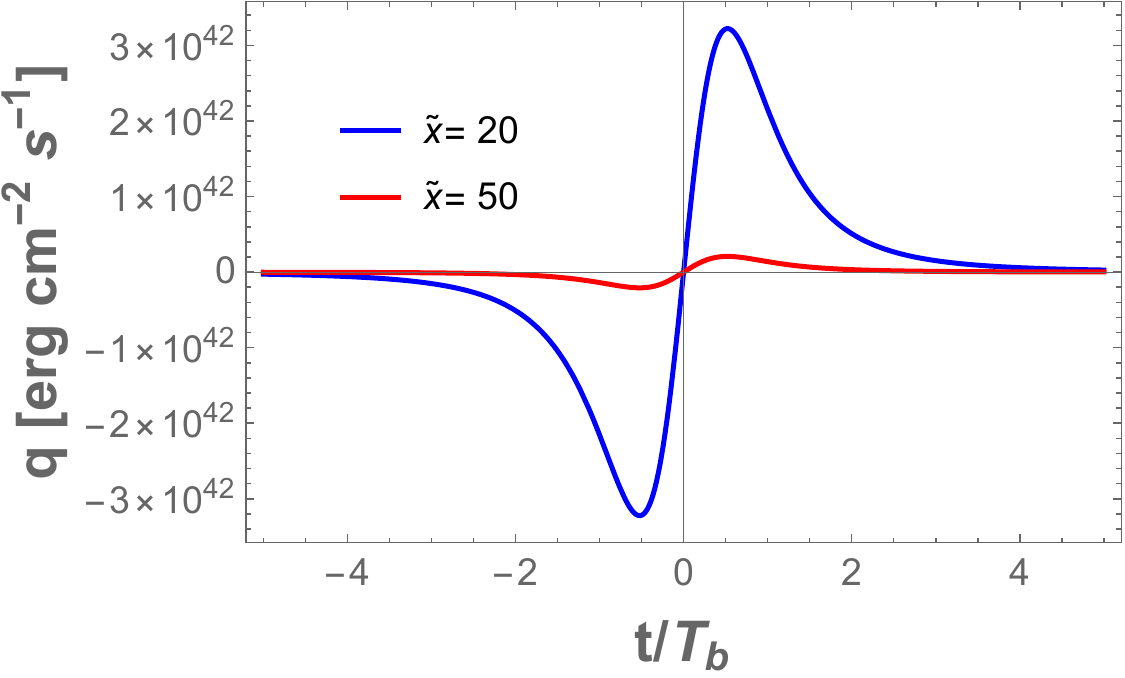}
    \caption{Heat flux as a function cosmic time for two fixed value of the radial coordinate: $\tilde{r} = 20\;r_{\rm g}$ and $\tilde{r} = 50\;r_{\rm g}$, corresponding to the Perez-Raia Neto wormhole solution. In this figure we adopt $T_b = 10^{-4}$ and $b_0 = 2 \; G M_{\odot} /c^2$. }
    \label{fig_q1}
\end{figure}

\begin{figure}
    \centering
    \includegraphics[width=1.\linewidth]{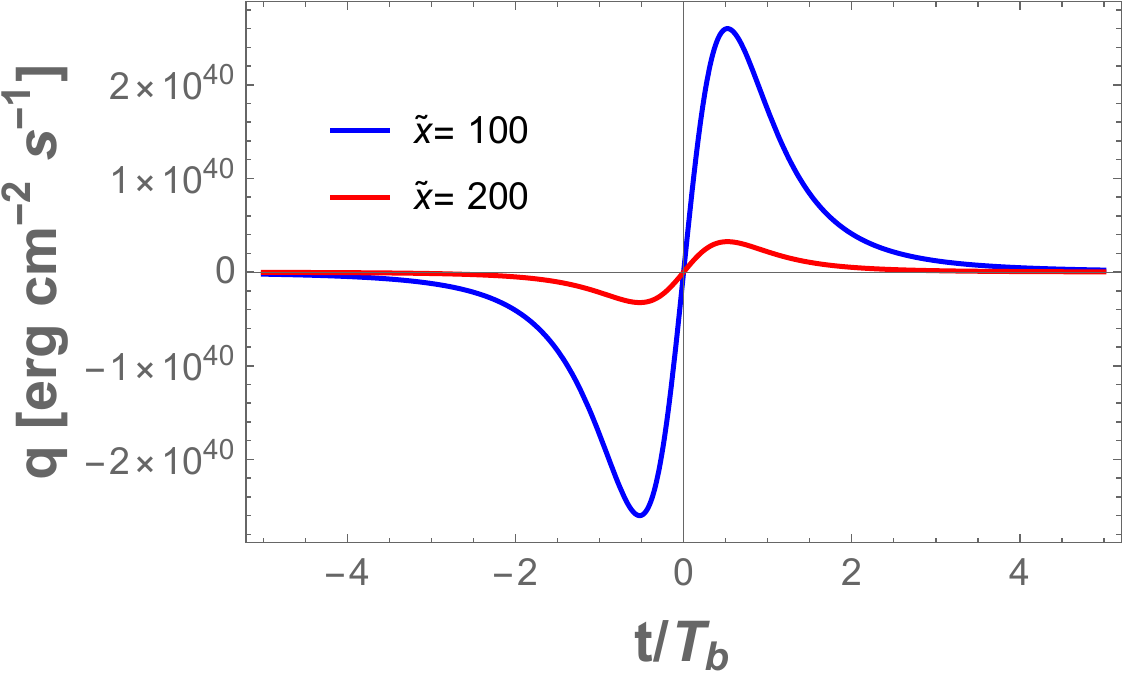}
    \caption{Heat flux as a function cosmic time for two fixed value of the radial coordinate: $\tilde{r} = 100\;r_{\rm g}$ and $\tilde{r} = 200\;r_{\rm g}$, corresponding to the Perez-Raia Neto wormhole solution. In this figure we adopt $T_b = 10^{-4}$ and $b_0 = 2 \; G M_{\odot} /c^2$. }
    \label{fig_q2}
\end{figure}

\begin{figure}
    \centering
    \includegraphics[width=1.\linewidth]{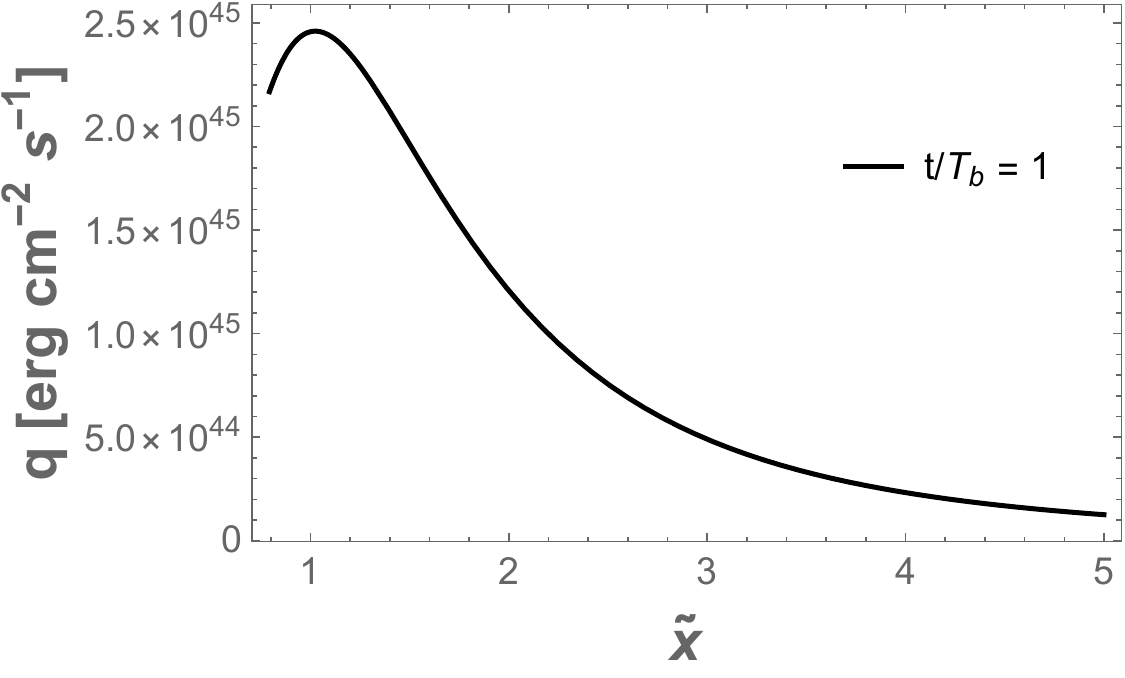}
    \caption{Heat flux as function of the isotropic radius $\tilde{x} = \tilde{r}/r_{\rm g}$ at $t/T_{\rm b} = 1$ corresponding to the Perez-Raia Neto wormhole solution.In this figure we adopt $T_b = 10^{-4}$ and $b_0 = 2 \; G M_{\odot} /c^2$. }
    \label{fig_qr1}
\end{figure}

\begin{figure}
    \centering
    \includegraphics[width=1.\linewidth]{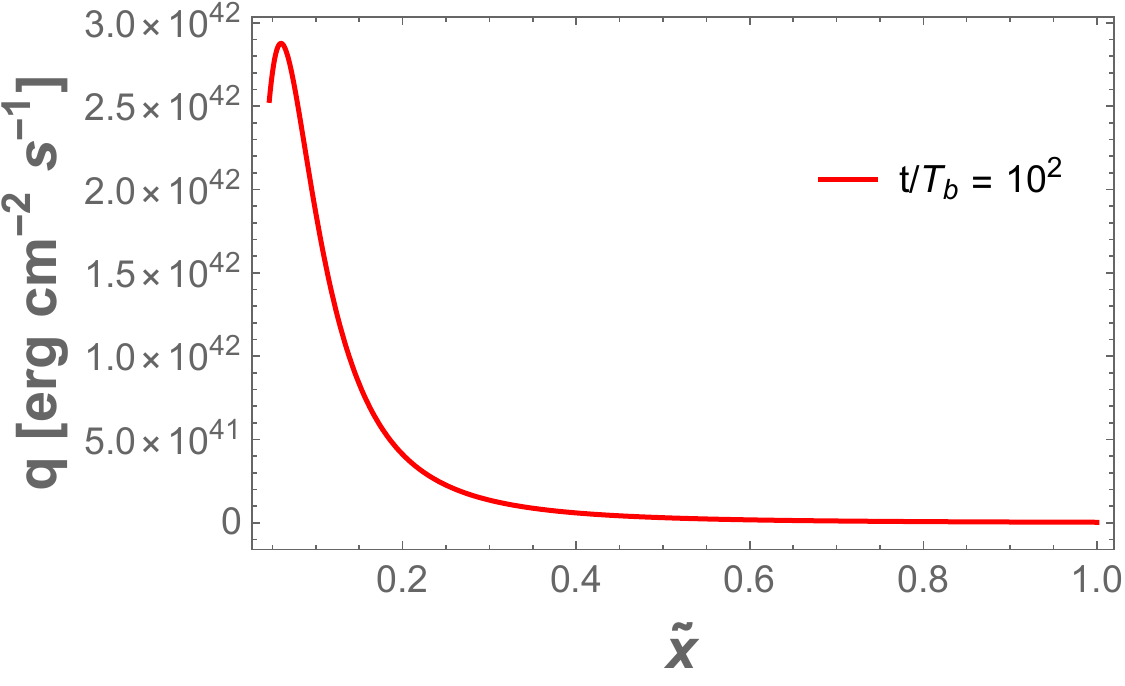}
    \caption{Heat flux as function of the isotropic radius $\tilde{x} = \tilde{r}/r_{\rm g}$ at $t/T_{\rm b} = 10^2$ corresponding to the Perez-Raia Neto wormhole solution. In this figure we adopt $T_b = 10^{-4}$ and $b_0 = 2 \; G M_{\odot} /c^2$. }
    \label{fig_qr2}
\end{figure}

\begin{figure}
    \centering
    \includegraphics[width=1.\linewidth]{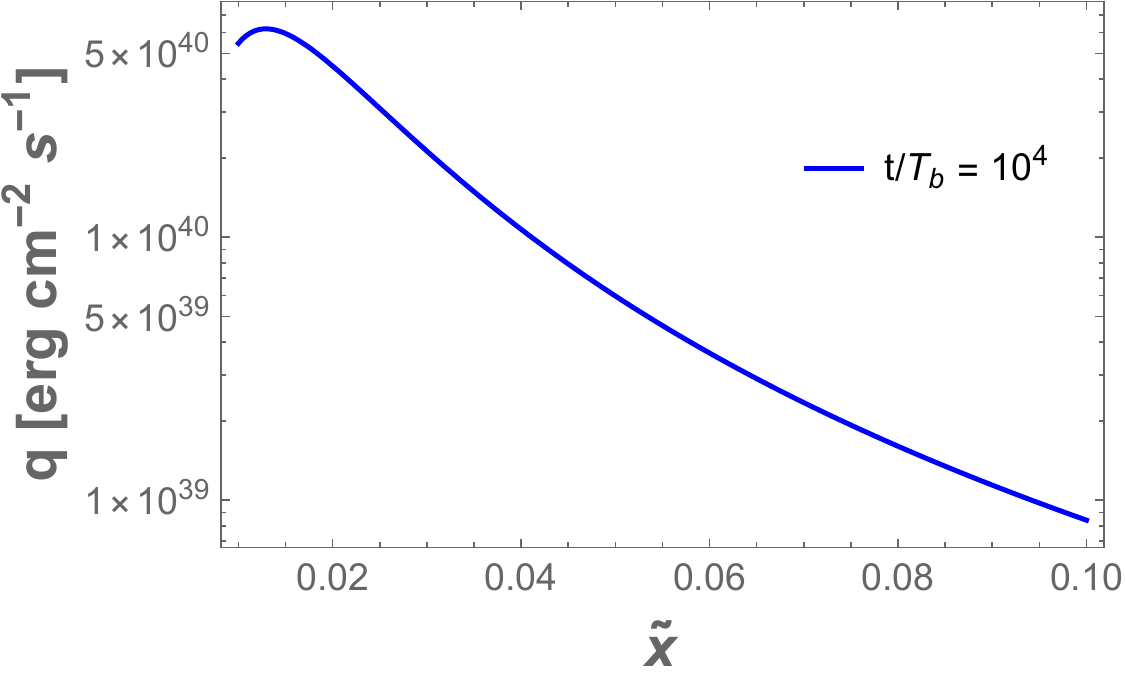}
    \caption{Heat flux as function of the isotropic radius $\tilde{x} = \tilde{r}/r_{\rm g}$ at $t/T_{\rm b} = 10^4$ corresponding to the Perez-Raia Neto wormhole solution. In this figure we adopt $T_b = 10^{-4}$ and $b_0 = 2 \; G M_{\odot} /c^2$. }
    \label{fig_qr3}
\end{figure}

One of the main differences between Kim's solution and the PRN solution is the type of energy-momentum tensor adopted. In the latter, there is a heat flux whose explicit form, assuming the scale factor \eqref{a-bounce}, is
\begin{equation}
q(\tilde{r},t)=\frac{128\: {a_{\rm b}}^4 {b_0}^2 \tilde{r}^5 t \:\sqrt[3]{\frac{t^2}{{T_{\rm b}}^2}+1}}{3 \pi  {T_{\rm b}}^2 \left[4 {a_{\rm b}}^2 \tilde{r}^2 \left(\frac{t^2}{{T_{\rm b}}^2}+1\right)^{2/3}+{b_0}^2\right]^4}.
\end{equation}

The heat flux is an odd function of the cosmic time. The dependence of $q$ with time at $\tilde{r}_{\rm th} =2 \; G M_{\odot} /c^2 $, and for different values of $\tilde{r}$ is shown in Figures \ref{fig_q1} and \ref{fig_q2}. For any value of $\tilde{r}$, the flux is in the direction of the throat before the bounce, and away from it afterwards. The heat flux goes rapidly to zero with increasing $t$.

Figures \ref{fig_qr1}, \ref{fig_qr2}, and \ref{fig_qr3} show $q$
as a function of $\tilde{r}$ for different values of $t/T_{\rm b}$. The plots start at the isotropic radius of the throat at the given cosmic time $r_{\rm thr}= b_0/(2 a(t_{*})$ with $t_{*}/T_{\rm b} = 1,\; 10^2 \; 10^4$. The heat flux decays rapidly with $t$ for any value of $r$, and has a maximum at $\tilde{r} = \sqrt{5/3} \tilde{r}_{\rm thr}$.\\

\section{Final remarks}

In this paper, we have shown that if the topology of the universe is multiply connected in the sense that it contains a wormhole, and there is a cosmological bounce from a contracting to an expanding phase of the cosmos, then both the dynamical wormhole solutions proposed by Kim and by P\'erez-Raia Neto survive the bounce, respectively. Let us emphasize that this is a non-trivial result, since 
the weak energy condition is violated at the non-singular model used in this work.

We find that, for the Kim solution, only wormholes with a  throat parameter $b_0$ smaller than $GM^{*}/c^2$, with $M^{*} \approx 22 \; M_{\odot}$  exist at all cosmic times. This is not the case for the other dynamical wormhole analyzed here, where no restriction must be imposed upon the solution for survival.


The survival of the wormhole opens some interesting possibilities that should be explored in the future. 
For instance, 
as recently shown by Ertola Urtubey et al. \cite{Milos2024EPJC}, the accretion of gas onto bounce-surviving wormholes in the early universe could produce jets. Such jets can also be a factor contributing to the re-ionization and the formation of the first astrophysical structures. 

Another interesting, albeit hypothetical, scenario would be that in which one or several wormholes connect the high-density contracting phase with the dark ages of the expanding one. The wormhole can then act as a Laval nozzle, accelerating the compressible cosmic fluid to supersonic speeds in the axial (thrust) direction by converting the thermal energy of the flow into kinetic energy. The result will be a very hot wind as shown by the GRMHD simulations recently performed by Combi et al. \cite{Combi2024PhRvD}. This wind could start the re-ionization of the universe at very early stages, even before the formation of the first stars. Moreover, the wind can play some role in the formation of the first stars and galaxies, pushing and perturbing the medium. 

Wormholes connecting the contracting to the expanding branch could be channels not only for matter flux, but also for gravitational waves. The specific properties of their contribution to the gravitational background could be used to constrain their potential existence and number. 

Let us remark that the throat of the Pérez-Raia Neto wormhole remains open regardless of the specific form of the scale factor (the radius of the throat in isotropic coordinates is $\tilde{r}_{\rm th} = b_0/(2 a(t))$). Thus, the survival of the wormhole is not restricted to a particular bouncing model: the wormhole survives provided that the background cosmological model is well-defined at all cosmic times. This is not the case for the Kim solution: this metric does not represent a wormhole embedded in a bouncing universe for $b_0 \gtrapprox 22 \; G M_{\odot}/c^2$.

Unfortunately, it is not possible to generalize the results obtained from the two specific models analyzed to all cosmological wormholes. However, the survival condition developed in this work can be applied to other cosmological wormholes in a bouncing universe. Whether a given wormhole persists through the bounce must be established on a case-by-case basis, by explicitly computing the location of the throat and verifying the flare-out condition.
In future work, we will explore other cosmological wormhole metrics, such as the so-called conformally expanding Morris-Thorne wormhole \citep{kar94,kar+06}, and determine whether they can persist throughout a cosmological bounce.

\section*{Acknowledgements}
D. P. acknowledges the support from CONICET under Grant No. PIP 0554 and AGENCIA I+D+i under Grant PICT- 2021-I-INVI-00387. G.E.R. acknowledges financial support from the State Agency for Research of the Spanish Ministry of Science and Innovation under grant PID2022-136828NB-C41/AEI/10.13039/501100011033/, and by
“ERDF A way of making Europe”, by the “European Union”. He also thanks support from PIP 0554 (CONICET) and from the \emph{Coordena\c c\~ao de Aperfei\c coamento de Pessoal de N\'ivel Superior-Brasil
(CAPES)-Codigo de Financiamento 001}.

\bibliography{apssamp_new}

@ARTICLE{Lopez-Corredoira2017FoPh,
       author = {{L{\'o}pez-Corredoira}, Mart{\'\i}n},
        title = "{Tests and Problems of the Standard Model in Cosmology}",
      journal = {Foundations of Physics},
         year = 2017,
        month = jun,
       volume = {47},
       number = {6},
        pages = {711-768},
          doi = {10.1007/s10701-017-0073-8},
archivePrefix = {arXiv},
       eprint = {1701.08720},
 primaryClass = {astro-ph.CO},
       adsurl = {https://ui.adsabs.harvard.edu/abs/2017FoPh...47..711L}
}

@ARTICLE{Tristram2024A&A,
       author = {{Tristram}, M. and {Banday}, A.~J. and {Douspis}, M. and {Garrido}, X. and {G{\'o}rski}, K.~M. and {Henrot-Versill{\'e}}, S. and {Hergt}, L.~T. and {Ili{\'c}}, S. and {Keskitalo}, R. and {Lagache}, G. and {Lawrence}, C.~R. and {Partridge}, B. and {Scott}, D.},
        title = "{Cosmological parameters derived from the final Planck data release (PR4)}",
      journal = {A\&A},
         year = 2024,
        month = feb,
       volume = {682},
          eid = {A37},
        pages = {A37},
          doi = {10.1051/0004-6361/202348015},
archivePrefix = {arXiv},
       eprint = {2309.10034},
 primaryClass = {astro-ph.CO},
       adsurl = {https://ui.adsabs.harvard.edu/abs/2024A&A...682A..37T}
}

@ARTICLE{Lu2022ApJ,
       author = {{Lu}, Jia and {Wang}, Lifan and {Chen}, Xingzhuo and {Rubin}, David and {Perlmutter}, Saul and {Baade}, Dietrich and {Mould}, Jeremy and {Vinko}, Jozsef and {Reg{\H{o}}s}, Enik{\H{o}} and {Koekemoer}, Anton M.},
        title = "{Constraints on Cosmological Parameters with a Sample of Type Ia Supernovae from JWST}",
      journal = {ApJ},
         year = 2022,
        month = dec,
       volume = {941},
       number = {1},
          eid = {71},
        pages = {71},
          doi = {10.3847/1538-4357/ac9f49},
archivePrefix = {arXiv},
       eprint = {2210.00746},
 primaryClass = {astro-ph.CO},
       adsurl = {https://ui.adsabs.harvard.edu/abs/2022ApJ...941...71L}
}

@ARTICLE{McCoy2015SHPMP,
       author = {{McCoy}, C.~D.},
        title = "{Does inflation solve the hot big bang model's fine-tuning problems?}",
      journal = {Studies in the History and Philosophy of Modern Physics},
         year = 2015,
        month = aug,
       volume = {51},
        pages = {23-36},
          doi = {10.1016/j.shpsb.2015.06.002},
       adsurl = {https://ui.adsabs.harvard.edu/abs/2015SHPMP..51...23M}
}

@ARTICLE{Milos2024EPJC,
       author = {{Ertola Urtubey}, Milos and {P{\'e}rez}, Daniela and {Romero}, Gustavo E.},
        title = "{Outgoing electromagnetic flux from rotating wormholes}",
      journal = {European Physical Journal C},
         year = 2024,
        month = nov,
       volume = {84},
       number = {11},
          eid = {1163},
        pages = {1163},
          doi = {10.1140/epjc/s10052-024-13563-2},
archivePrefix = {arXiv},
       eprint = {2411.13474},
 primaryClass = {gr-qc},
       adsurl = {https://ui.adsabs.harvard.edu/abs/2024EPJC...84.1163E}
}

@ARTICLE{Ijjas+2013PhLB,
       author = {{Ijjas}, Anna and {Steinhardt}, Paul J. and {Loeb}, Abraham},
        title = "{Inflationary paradigm in trouble after Planck2013}",
      journal = {Physics Letters B},
         year = 2013,
        month = jun,
       volume = {723},
       number = {4-5},
        pages = {261-266},
          doi = {10.1016/j.physletb.2013.05.023},
archivePrefix = {arXiv},
       eprint = {1304.2785},
 primaryClass = {astro-ph.CO},
       adsurl = {https://ui.adsabs.harvard.edu/abs/2013PhLB..723..261I}
}

@ARTICLE{Ijjas+2014PhLB,
       author = {{Ijjas}, Anna and {Steinhardt}, Paul J. and {Loeb}, Abraham},
        title = "{Inflationary schism}",
      journal = {Physics Letters B},
         year = 2014,
        month = sep,
       volume = {736},
        pages = {142-146},
          doi = {10.1016/j.physletb.2014.07.012},
archivePrefix = {arXiv},
       eprint = {1402.6980},
 primaryClass = {astro-ph.CO},
       adsurl = {https://ui.adsabs.harvard.edu/abs/2014PhLB..736..142I}
}

@article{Earman1999L,
	author = {John Earman and Jesus Mosterin},
	doi = {10.1086/392675},
	journal = {Philosophy of Science},
	number = {1},
	pages = {1--49},
	publisher = {University of Chicago Press},
	title = {A Critical Look at Inflationary Cosmology},
	volume = {66},
	year = {1999}
}

@INCOLLECTION{Brandenberger2008LNP,
       author = {{Brandenberger}, Robert H.},
        title = "{Conceptual Problems of Inflationary Cosmology and a New Approach to Cosmological Structure Formation}",
    booktitle = {Inflationary Cosmology},
         year = 2008,
       editor = {{Lemoine}, Martin and {Martin}, Jerome and {Peter}, Patrick},
       volume = {738},
        pages = {393},
          doi = {10.1007/978-3-540-74353-8_11},
       adsurl = {https://ui.adsabs.harvard.edu/abs/2008LNP...738..393B}
}

@ARTICLE{Agullo+2021,
       author = {{Agullo}, Ivan and {Kranas}, Dimitrios and {Sreenath}, V.},
        title = "{Large scale anomalies in the CMB and non-Gaussianity in bouncing cosmologies}",
      journal = {Classical and Quantum Gravity},
         year = 2021,
        month = mar,
       volume = {38},
       number = {6},
          eid = {065010},
        pages = {065010},
          doi = {10.1088/1361-6382/abc521},
archivePrefix = {arXiv},
       eprint = {2006.09605},
 primaryClass = {astro-ph.CO},
       adsurl = {https://ui.adsabs.harvard.edu/abs/2021CQGra..38f5010A}
}

@ARTICLE{Borde1994,
       author = {{Borde}, Arvind},
        title = "{Topology Change in Classical General Relativity}",
      journal = {arXiv e-prints},
         year = 1994,
        month = jun,
          eid = {gr-qc/9406053},
        pages = {},
          doi = {10.48550/arXiv.gr-qc/9406053},
archivePrefix = {arXiv},
       eprint = {},
 primaryClass = {gr-qc},
       adsurl = {https://ui.adsabs.harvard.edu/abs/1994gr.qc.....6053B}
}

@ARTICLE{Joshi1987PhLA,
       author = {{Joshi}, P.~S. and {Saraykar}, R.~V.},
        title = "{Cosmic censorship and topology change in general relativity}",
      journal = {Physics Letters A},
         year = 1987,
        month = feb,
       volume = {120},
       number = {3},
        pages = {111-114},
          doi = {10.1016/0375-9601(87)90708-0},
       adsurl = {https://ui.adsabs.harvard.edu/abs/1987PhLA..120..111J}
}

@ARTICLE{Hayward1999IJMPD,
       author = {{Hayward}, Sean A.},
        title = "{Dynamic Wormholes}",
      journal = {International Journal of Modern Physics D},
         year = 1999,
        month = jan,
       volume = {8},
       number = {3},
        pages = {373-382},
          doi = {10.1142/S0218271899000286},
archivePrefix = {arXiv},
       eprint = {gr-qc/9805019},
 primaryClass = {gr-qc},
       adsurl = {https://ui.adsabs.harvard.edu/abs/1999IJMPD...8..373H}
}

@ARTICLE{Tomikawa2015PhRvD,
       author = {{Tomikawa}, Yoshimune and {Izumi}, Keisuke and {Shiromizu}, Tetsuya},
        title = "{New definition of a wormhole throat}",
      journal = {Phys. Rev. D},
         year = 2015,
        month = may,
       volume = {91},
       number = {10},
          eid = {104008},
        pages = {104008},
          doi = {10.1103/PhysRevD.91.104008},
archivePrefix = {arXiv},
       eprint = {1503.01926},
 primaryClass = {gr-qc},
       adsurl = {https://ui.adsabs.harvard.edu/abs/2015PhRvD..91j4008T}
}

@ARTICLE{Kim2020IJMPD,
       author = {{Kim}, Sung-Won},
        title = "{Evolution of cosmological horizons of wormhole cosmology}",
      journal = {International Journal of Modern Physics D},
         year = 2020,
        month = jan,
       volume = {29},
       number = {12},
          eid = {2050079},
        pages = {2050079},
          doi = {10.1142/S0218271820500790},
       adsurl = {https://ui.adsabs.harvard.edu/abs/2020IJMPD..2950079K}
}

@ARTICLE{Kirilov2016IJMPD,
       author = {{Kirillov}, A.~A. and {Savelova}, E.~P.},
        title = "{Cosmological wormholes}",
      journal = {International Journal of Modern Physics D},
         year = 2016,
        month = may,
       volume = {25},
       number = {6},
          eid = {1650075},
        pages = {1650075},
          doi = {10.1142/S0218271816500759},
archivePrefix = {arXiv},
       eprint = {1512.01450},
 primaryClass = {gr-qc},
       adsurl = {https://ui.adsabs.harvard.edu/abs/2016IJMPD..2550075K}
}

@ARTICLE{Lobo2007,
       author = {{Lobo}, Francisco S.~N.},
        title = "{Exotic solutions in General Relativity: Traversable wormholes and 'warp drive' spacetimes}",
      journal = {arXiv e-prints},
         year = 2007,
        month = oct,
          eid = {arXiv:0710.4474},
        pages = {arXiv:0710.4474},
          doi = {10.48550/arXiv.0710.4474},
archivePrefix = {arXiv},
       eprint = {0710.4474},
 primaryClass = {gr-qc},
       adsurl = {https://ui.adsabs.harvard.edu/abs/2007arXiv0710.4474L}
}

@BOOK{Visser1996book,
       author = {{Visser}, Matt},
        title = "{Lorentzian Wormholes}",
         year = 1996,
        publisher= "{Springer-Verlag}",
       adsurl = {https://ui.adsabs.harvard.edu/abs/1996lowo.book.....V}
}

@ARTICLE{Safonova2001PhRvD,
       author = {{Safonova}, Margarita and {Torres}, Diego F. and {Romero}, Gustavo E.},
        title = "{Microlensing by natural wormholes: Theory and simulations}",
      journal = {Phys. Rev. D},
         year = 2001,
        month = dec,
       volume = {65},
       number = {2},
          eid = {023001},
        pages = {023001},
          doi = {10.1103/PhysRevD.65.023001},
archivePrefix = {arXiv},
       eprint = {gr-qc/0105070},
 primaryClass = {gr-qc},
       adsurl = {https://ui.adsabs.harvard.edu/abs/2001PhRvD..65b3001S}
}

@ARTICLE{Torres1998PhRvD,
       author = {{Torres}, Diego F. and {Romero}, Gustavo E. and {Anchordoqui}, Luis A.},
        title = "{Might some gamma ray bursts be an observable signature of natural wormholes?}",
      journal = {Phys. Rev. D},
         year = 1998,
        month = dec,
       volume = {58},
       number = {12},
          eid = {123001},
        pages = {123001},
          doi = {10.1103/PhysRevD.58.123001},
archivePrefix = {arXiv},
       eprint = {astro-ph/9802106},
 primaryClass = {astro-ph},
       adsurl = {https://ui.adsabs.harvard.edu/abs/1998PhRvD..58l3001T}
}

@ARTICLE{Bambi2021Univ,
       author = {{Bambi}, Cosimo and {Stojkovic}, Dejan},
        title = "{Astrophysical Wormholes}",
      journal = {Universe},
         year = 2021,
        month = may,
       volume = {7},
       number = {5},
          eid = {136},
        pages = {136},
          doi = {10.3390/universe7050136},
archivePrefix = {arXiv},
       eprint = {2105.00881},
 primaryClass = {gr-qc},
       adsurl = {https://ui.adsabs.harvard.edu/abs/2021Univ....7..136B}
}

@ARTICLE{Combi2024PhRvD,
       author = {{Combi}, Luciano and {Yang}, Huan and {Gutierrez}, Eduardo and {Noble}, Scott C. and {Romero}, Gustavo E. and {Campanelli}, Manuela},
        title = "{General relativistic magnetohydrodynamical simulations of accretion flows through traversable wormholes}",
      journal = {Phys. Rev. D},
         year = 2024,
        month = may,
       volume = {109},
       number = {10},
          eid = {103034},
        pages = {103034},
          doi = {10.1103/PhysRevD.109.103034},
archivePrefix = {arXiv},
       eprint = {2405.06900},
 primaryClass = {astro-ph.HE},
       adsurl = {https://ui.adsabs.harvard.edu/abs/2024PhRvD.109j3034C}
}

@ARTICLE{Perez2022PhRvD,
       author = {{P{\'e}rez}, Daniela and {Romero}, Gustavo E.},
        title = "{Survival of black holes through a cosmological bounce}",
      journal = {Phys. Rev. D},
         year = 2022,
        month = may,
       volume = {105},
       number = {10},
          eid = {104047},
        pages = {104047},
          doi = {10.1103/PhysRevD.105.104047},
archivePrefix = {arXiv},
       eprint = {2205.10333},
 primaryClass = {gr-qc},
       adsurl = {https://ui.adsabs.harvard.edu/abs/2022PhRvD.105j4047P}
}

@ARTICLE{Corman2022JCAP,
       author = {{Corman}, Maxence and {East}, William E. and {Ripley}, Justin L.},
        title = "{Evolution of black holes through a nonsingular cosmological bounce}",
      journal = {JCAP},
         year = 2022,
        month = sep,
       volume = {2022},
       number = {9},
          eid = {063},
        pages = {063},
          doi = {10.1088/1475-7516/2022/09/063},
archivePrefix = {arXiv},
       eprint = {2206.08466},
 primaryClass = {gr-qc},
       adsurl = {https://ui.adsabs.harvard.edu/abs/2022JCAP...09..063C}
}

@ARTICLE{Perez2021PhRvDP,
       author = {{P{\'e}rez}, Daniela and {Bergliaffa}, Santiago E. Perez and {Romero}, Gustavo E.},
        title = "{Dynamical black hole in a bouncing universe}",
      journal = {Phys. Rev. D},
         year = 2021,
        month = mar,
       volume = {103},
       number = {6},
          eid = {064019},
        pages = {064019},
          doi = {10.1103/PhysRevD.103.064019},
archivePrefix = {arXiv},
       eprint = {2103.00108},
 primaryClass = {gr-qc},
       adsurl = {https://ui.adsabs.harvard.edu/abs/2021PhRvD.103f4019P}
}

@ARTICLE{Ijjas2024JCAP,
       author = {{Ijjas}, Anna and {Steinhardt}, Paul J. and {Garfinkle}, David and {Cook}, William G.},
        title = "{Smoothing and flattening the universe through slow contraction versus inflation}",
      journal = {JCAP},
         year = 2024,
        month = jul,
       volume = {2024},
       number = {7},
          eid = {077},
        pages = {077},
          doi = {10.1088/1475-7516/2024/07/077},
archivePrefix = {arXiv},
       eprint = {2404.00867},
 primaryClass = {gr-qc},
       adsurl = {https://ui.adsabs.harvard.edu/abs/2024JCAP...07..077I}
}

@ARTICLE{Ijjas2016,
       author = {{Ijjas}, Anna and {Steinhardt}, Paul J.},
        title = "{Classically Stable Nonsingular Cosmological Bounces}",
      journal = {Phys.\ Rev.\ Lett.},
         year = 2016,
        month = sep,
       volume = {117},
       number = {12},
          eid = {121304},
        pages = {121304},
          doi = {10.1103/PhysRevLett.117.121304},
archivePrefix = {arXiv},
       eprint = {1606.08880},
 primaryClass = {gr-qc},
       adsurl = {https://ui.adsabs.harvard.edu/abs/2016PhRvL.117l1304I}
}

@INPROCEEDINGS{Pinto-Neto2011,
       author = {{Pinto-Neto}, Nelson},
        title = "{The Bounce Confronts the Big-Bang}",
    booktitle = {The Sun, the Stars, the Universe and General Relativity},
         year = 2011,
       editor = {{Perez Berliaffa}, S.~E. and {Novello}, M. and {Ruffini}, R.},
        month = jan,
        pages = {227-240},
       adsurl = {https://ui.adsabs.harvard.edu/abs/2011ssug.conf..227P}
}

@ARTICLE{Pinto-Neto2014AN,
       author = {{Pinto-Neto}, N.},
        title = "{Bouncing models and inflation}",
      journal = {Astronomische Nachrichten},
         year = 2014,
        month = sep,
       volume = {335},
       number = {6-7},
        pages = {727},
          doi = {10.1002/asna.201412100},
       adsurl = {https://ui.adsabs.harvard.edu/abs/2014AN....335..727P}
}

@ARTICLE{Pinto-Neto2021,
       author = {{Pinto-Neto}, Nelson},
        title = "{Bouncing Quantum Cosmology}",
      journal = {Universe},
         year = 2021,
        month = apr,
       volume = {7},
       number = {4},
          eid = {110},
        pages = {110},
          doi = {10.3390/universe7040110},
       adsurl = {https://ui.adsabs.harvard.edu/abs/2021Univ....7..110P}
}

@PREAMBLE{
 "\providecommand{\noopsort}[1]{}" 
 # "\providecommand{\singleletter}[1]{#1}%" 
}

@article{hoc+98a,
  title = {Null Energy Condition in Dynamic Wormholes},
  author = {Hochberg, David and Visser, Matt},
  journal = {Phys. Rev. Lett.},
  volume = {81},
  issue = {4},
  pages = {746--749},
  numpages = {0},
  year = {1998},
  month = {Jul},
  publisher = {American Physical Society},
  doi = {10.1103/PhysRevLett.81.746},
  url = {https://link.aps.org/doi/10.1103/PhysRevLett.81.746}
}

@ARTICLE{hoc+98b,
       author = {{Hochberg}, David and {Visser}, Matt},
        title = "{Dynamic wormholes, antitrapped surfaces, and energy conditions}",
      journal = {Physical Review D},
         year = 1998,
        month = aug,
       volume = {58},
       number = {4},
          eid = {044021},
        pages = {044021},
          doi = {10.1103/PhysRevD.58.044021},
archivePrefix = {arXiv},
       eprint = {gr-qc/9802046},
 primaryClass = {gr-qc},
       adsurl = {https://ui.adsabs.harvard.edu/abs/1998PhRvD..58d4021H}
}

@ARTICLE{kar94,
       author = {{Kar}, Sayan},
        title = "{Evolving wormholes and the weak energy condition}",
      journal = {Physical Review D},
         year = 1994,
        month = jan,
       volume = {49},
       number = {2},
        pages = {862-865},
          doi = {10.1103/PhysRevD.49.862},
       adsurl = {https://ui.adsabs.harvard.edu/abs/1994PhRvD..49..862K}
}

@ARTICLE{kar+06,
       author = {{Kar}, Sayan and {Sahdev}, Deshdeep},
        title = "{Evolving Lorentzian wormholes}",
      journal = {Physical Review D},
         year = 1996,
        month = jan,
       volume = {53},
       number = {2},
        pages = {722-730},
          doi = {10.1103/PhysRevD.53.722},
archivePrefix = {arXiv},
       eprint = {gr-qc/9506094},
 primaryClass = {gr-qc},
       adsurl = {https://ui.adsabs.harvard.edu/abs/1996PhRvD..53..722K}
}

@ARTICLE{kim96,
       author = {{Kim}, Sung-Won},
        title = "{Cosmological model with a traversable wormhole}",
      journal = {Physical Review D},
         year = 1996,
        month = jun,
       volume = {53},
       number = {12},
        pages = {6889-6892},
          doi = {10.1103/PhysRevD.53.6889},
       adsurl = {https://ui.adsabs.harvard.edu/abs/1996PhRvD..53.6889K}
}

@ARTICLE{kim18,
       author = {{Kim}, Sung-Won},
        title = "{The cosmological model with a wormhole and Hawking temperature near apparent horizon}",
      journal = {Physics Letters B},
         year = 2018,
        month = may,
       volume = {780},
        pages = {174-180},
          doi = {10.1016/j.physletb.2018.03.005},
archivePrefix = {arXiv},
       eprint = {1801.07989},
 primaryClass = {gr-qc},
       adsurl = {https://ui.adsabs.harvard.edu/abs/2018PhLB..780..174K}
}

@ARTICLE{kim20,
       author = {{Kim}, Sung-Won},
        title = "{Evolution of cosmological horizons of wormhole cosmology}",
      journal = {International Journal of Modern Physics D},
         year = 2020,
        month = jan,
       volume = {29},
       number = {12},
          eid = {2050079},
        pages = {2050079},
          doi = {10.1142/S0218271820500790},
       adsurl = {https://ui.adsabs.harvard.edu/abs/2020IJMPD..2950079K}
}

@ARTICLE{mae+22,
       author = {{Maeda}, Hideki and {Harada}, Tomohiro},
        title = "{Criteria for energy conditions}",
      journal = {Classical and Quantum Gravity},
         year = 2022,
        month = oct,
       volume = {39},
       number = {19},
          eid = {195002},
        pages = {195002},
          doi = {10.1088/1361-6382/ac8861},
archivePrefix = {arXiv},
       eprint = {2205.12993},
 primaryClass = {gr-qc},
       adsurl = {https://ui.adsabs.harvard.edu/abs/2022CQGra..39s5002M}
}

@ARTICLE{mae+09,
       author = {{Maeda}, Hideki and {Harada}, Tomohiro and {Carr}, B.~J.},
        title = "{Cosmological wormholes}",
      journal = {Physical Review D},
         year = 2009,
        month = feb,
       volume = {79},
       number = {4},
          eid = {044034},
        pages = {044034},
          doi = {10.1103/PhysRevD.79.044034},
archivePrefix = {arXiv},
       eprint = {0901.1153},
 primaryClass = {gr-qc},
       adsurl = {https://ui.adsabs.harvard.edu/abs/2009PhRvD..79d4034M}
}

@ARTICLE{mcn+21,
       author = {{McNutt}, D.~D. and {Julius}, W. and {Gorban}, M. and {Mattingly}, B. and {Brown}, P. and {Cleaver}, G.},
        title = "{Geometric surfaces: An invariant characterization of spherically symmetric black hole horizons and wormhole throats}",
      journal = {Physical Review D},
         year = 2021,
        month = jun,
       volume = {103},
       number = {12},
          eid = {124024},
        pages = {124024},
          doi = {10.1103/PhysRevD.103.124024},
archivePrefix = {arXiv},
       eprint = {2104.08935},
 primaryClass = {gr-qc},
       adsurl = {https://ui.adsabs.harvard.edu/abs/2021PhRvD.103l4024M}
}

@ARTICLE{mor+88,
       author = {{Morris}, Michael S. and {Thorne}, Kip S.},
        title = "{Wormholes in spacetime and their use for interstellar travel: A tool for teaching general relativity}",
      journal = {American Journal of Physics},
         year = 1988,
        month = may,
       volume = {56},
       number = {5},
        pages = {395-412},
          doi = {10.1119/1.15620},
       adsurl = {https://ui.adsabs.harvard.edu/abs/1988AmJPh..56..395M}
}

@ARTICLE{per+23,
       author = {{P{\'e}rez}, Daniela and {Neto}, M{\'a}rio Raia},
        title = "{A new solution for a generalized cosmological wormhole}",
      journal = {European Physical Journal C},
         year = 2023,
        month = dec,
       volume = {83},
       number = {12},
          eid = {1127},
        pages = {1127},
          doi = {10.1140/epjc/s10052-023-12316-x},
archivePrefix = {arXiv},
       eprint = {2312.07736},
 primaryClass = {gr-qc},
       adsurl = {https://ui.adsabs.harvard.edu/abs/2023EPJC...83.1127P}
}

@BOOK{far15,
       author = {{Faraoni}, Valerio},
        title = "{Cosmological and Black Hole Apparent Horizons}",
         year = 2015,
       volume = {907},
          doi = {10.1007/978-3-319-19240-6},
       adsurl = {https://ui.adsabs.harvard.edu/abs/2015LNP...907.....F}
}

@BOOK{Hawking-Ellis-book,
       author = {{Hawking}, S.~W. and {Ellis}, G.~F.~R.},
        title = "{The large-scale structure of space-time.}",
         year = 1973,
       adsurl = {https://ui.adsabs.harvard.edu/abs/1973lsss.book.....H}
}

@ARTICLE{kor+25,
       author = {{Kord Zangeneh}, Mahdi and {Lobo}, Francisco S.~N.},
        title = "{Evolving Wormholes in a Cosmological Background}",
      journal = {Universe},
         year = 2025,
        month = jul,
       volume = {11},
       number = {7},
          eid = {236},
        pages = {236},
          doi = {10.3390/universe11070236},
archivePrefix = {arXiv},
       eprint = {2507.14750},
 primaryClass = {gr-qc},
       adsurl = {https://ui.adsabs.harvard.edu/abs/2025Univ...11..236K}
}

@ARTICLE{pet+07,
       author = {{Peter}, Patrick and {Pinho}, Emanuel J.~C. and {Pinto-Neto}, Nelson},
        title = "{Noninflationary model with scale invariant cosmological perturbations}",
      journal = prd,
         year = 2007,
        month = jan,
       volume = {75},
       number = {2},
          eid = {023516},
        pages = {023516},
          doi = {10.1103/PhysRevD.75.023516},
archivePrefix = {arXiv},
       eprint = {hep-th/0610205},
 primaryClass = {hep-th},
       adsurl = {https://ui.adsabs.harvard.edu/abs/2007PhRvD..75b3516P}
}

@article{Ijjas:2016tpn,
    author = "Ijjas, Anna and Steinhardt, Paul J.",
    archivePrefix = "arXiv",
    doi = "10.1103/PhysRevLett.117.121304",
    eprint = "1606.08880",
    journal = prl,
    number = "12",
    pages = "121304",
    primaryClass = "gr-qc",
    title = "{Classically stable nonsingular cosmological bounces}",
    volume = "117",
    year = "2016"
}

@article{Peter:2008qz,
    author = "Peter, Patrick and Pinto-Neto, Nelson",
    archivePrefix = "arXiv",
    doi = "10.1103/PhysRevD.78.063506",
    eprint = "0809.2022",
    journal = prd,
    pages = "063506",
    primaryClass = "gr-qc",
    title = "{Cosmology without inflation}",
    volume = "78",
    year = "2008"
}

@article{Almeida:2018xvj,
    author = "Almeida, C.R. and Bergeron, H. and Gazeau, J.-P. and Scardua, A.C.",
    doi = "10.1016/j.aop.2018.03.010",
    journal = "Annals Phys.",
    pages = "206--228",
    title = "{Three examples of quantum dynamics on the half-line with smooth bouncing}",
    volume = "392",
    year = "2018",
    eprint = "1708.06422",
    primaryClass = "quant-ph"
}

@article{Falciano:2008gt,
    author = "Falciano, Felipe T. and Lilley, Marc and Peter, Patrick",
    title = "{A Classical bounce: Constraints and consequences}",
    eprint = "0802.1196",
    archivePrefix = "arXiv",
    primaryClass = "gr-qc",
    doi = "10.1103/PhysRevD.77.083513",
    journal = prd,
    volume = "77",
    pages = "083513",
    year = "2008"
}

@article{Galkina:2019pir,
    author = "Galkina, O. and Fabris, J.C. and Falciano, F.T. and Pinto-Neto, N.",
    archivePrefix = "arXiv",
    doi = "10.1134/S0021364019200013",
    eprint = "1908.04258",
    journal = "JETP Lett.",
    number = "8",
    pages = "523--528",
    primaryClass = "gr-qc",
    title = "{Regular Bouncing Solutions, Energy Conditions, and the Brans---Dicke Theory}",
    volume = "110",
    year = "2019"
}

@article{Frion2020,
    author = "Frion, E. and Pinto-Neto, N. and Vitenti, S.D.P. and Perez Bergliaffa, S.E.",
    title = "{Primordial Magnetogenesis in a Bouncing Universe}",
    eprint = "2004.07269",
    archivePrefix = "arXiv",
    primaryClass = "gr-qc",
    doi = "10.1103/PhysRevD.101.103503",
    journal = "Phys. Rev. D",
    volume = "101",
    number = "10",
    pages = "103503",
    year = "2020"
}

@article{Novello2008,
      author         = "Novello, M. and Bergliaffa, S. E. Perez",
      title          = "{Bouncing Cosmologies}",
      journal        = "Phys. Rept.",
      volume         = "463",
      year           = "2008",
      pages          = "127-213",
      doi            = "10.1016/j.physrep.2008.04.006",
      eprint         = "0802.1634",
      archivePrefix  = "arXiv",
      primaryClass   = "astro-ph",
      SLACcitation   = "%%CITATION = ARXIV:0802.1634;%%"
}

@article{brandenburg2005astrophysical,
	title={Astrophysical magnetic fields and nonlinear dynamo theory},
	author={Brandenburg, Axel and Subramanian, Kandaswamy},
	journal={Physics Reports},
	volume={417},
	number={1-4},
	pages={1--209},
	year={2005},
	publisher={Elsevier}
}

@article{Bacalhau:2017hja,
    author = "Bacalhau, Anna Paula and Pinto-Neto, Nelson and Dias Pinto Vitenti, Sandro",
    archivePrefix = "arXiv",
    doi = "10.1103/PhysRevD.97.083517",
    eprint = "1706.08830",
    journal = "Phys.\ Rev.\ D",
    number = "8",
    pages = "083517",
    primaryClass = "gr-qc",
    title = "{Consistent Scalar and Tensor Perturbation Power Spectra in Single Fluid Matter Bounce with Dark Energy Era}",
    volume = "97",
    year = "2018"
}


\section*{Declarations}

Some journals require declarations to be submitted in a standardized format. Please check the Instructions for Authors of the journal to which you are submitting to see if you need to complete this section. If yes, your manuscript must contain the following sections under the heading `Declarations':

\begin{itemize}
\item Funding
\item Conflict of interest/Competing interests (check journal-specific guidelines for which heading to use)
\item Ethics approval and consent to participate
\item Consent for publication
\item Data availability 
\item Materials availability
\item Code availability 
\item Author contribution
\end{itemize}

\noindent
If any of the sections are not relevant to your manuscript, please include the heading and write `Not applicable' for that section. 







\end{document}